\RequirePackage[T1]{fontenc}
\documentclass[12pt]{article}
\pdfoutput=1 
\usepackage[height=8.85in,width=6.45in]{geometry}

\usepackage[utf8]{inputenc}
\usepackage{amsmath}
\usepackage{amssymb}
\usepackage{mathtools}
\numberwithin{equation}{section}
\usepackage{slashed}
\usepackage{braket}
\usepackage[svgnames]{xcolor}
\usepackage[colorlinks,linktocpage=true,citecolor=DarkGreen,linkcolor=FireBrick]{hyperref}
\usepackage{cite}
\usepackage{graphicx}
\usepackage{tikz}
\usepackage{tikz-cd}
\usepackage{times}
\usepackage[scaled]{couriers}
\usepackage{bm}
\usepackage{subfig}
\usepackage{mathrsfs}
\usepackage{amsthm}

\usepackage{xcolor}
\usepackage{mdframed}

\renewenvironment{figure}[1][]{
  \begin{originalfigure}[#1]
    \begin{mdframed}[linecolor=black!0,backgroundcolor=black!1]
}{
    \end{mdframed}
  \end{originalfigure}
}

\theoremstyle{plain}

\theoremstyle{definition}

\numberwithin{thm}{section}

\def\d{{\rm d}}
\def\i{{\mathsf i}}

\DeclareMathOperator{\tr}{tr}

\def\Im{\mathop{\mathrm{Im}}}
\def\diag{\mathop{\rm diag}\nolimits}

\def\Hom{\mathop{\mathrm{Hom}}}
\def\Ext{\mathop{\mathrm{Ext}}}
\usepackage{slashed}

\def\Ker{\mathop{\mathrm{Ker}}}
\def\Kerp{\mathop{ \widetilde{\mathrm{Ker}}}}
\def\index{\mathop{\mathrm{Index}}}
\def\indexp{\mathop{\widetilde{\mathrm{Index}}}}

\def\cA{{\cal A}}
\def\cB{{\cal B}}
\def\cC{{\cal C}}
\def\cD{{\cal D}}

\def\cH{{\cal H}}
\def\cI{{\cal I}}
\def\cJ{{\cal J}}

\def\cL{{\cal L}}

\def\cN{{\cal N}}

\def\cP{{\cal P}}

\def\cS{{\cal S}}
\def\cT{{\cal T}}

\def\cW{{\cal W}}
\def\cX{{\cal X}}
\def\cY{{\cal Y}}
\def\cZ{{\cal Z}}

\def\bA{{\mathbb A}}

\def\bQ{{\mathbb Q}}
\def\bR{{\mathbb R}}

\def\bZ{{\mathbb Z}}

\def\sM{{\mathsf M}}

\def\sT{{\mathsf T}}

\def\si{{\mathsf i}}

\def\sm{{\mathsf m}}

\def\so{{\mathsf o}}

\def\U{\mathrm{U}}
\def\SU{\mathrm{SU}}
\def\O{\mathrm{O}}
\def\SO{\mathrm{SO}}

\def\Spin{\mathrm{Spin}}
\def\Pin{\mathrm{Pin}}
\def\SL{\mathrm{SL}}

\def\so{\mathfrak{so}}

\def\spin{\mathrm{spin}}
\def\pin{\mathrm{pin}}
\def\beq#1\eeq{\begin{align}#1\end{align}}

\def\pt{\mathrm{pt}}
\def\IOspin{I\Omega^\textrm{spin}}
\def\IO{I\Omega}

\begin{document}

\begin{titlepage}

\begin{flushright}
TU-1163
\end{flushright}

\vskip 3cm

\begin{center}

{\Large \bfseries Heterotic global anomalies and torsion Witten index}

\vskip 1cm
Kazuya Yonekura
\vskip 1cm

\begin{tabular}{ll}
Department of Physics, Tohoku University, Sendai 980-8578, Japan
\end{tabular}

\vskip 1cm

\end{center}

\noindent
We study the structure of anomalies in general heterotic string theories by considering
general 2-dimensional $\cN=(0,1)$ supersymmetric quantum field theories (SQFTs), 
without assuming conformal invariance nor the correct central charges. 
First we generalize the precise notion of the $B$-field introduced by Witten. 
Then we express the target space anomalies by invariants of SQFTs.
Perturbative anomalies correspond to the Witten index of some class of SQFTs, 
while global anomalies correspond to a torsion version of the Witten index.
The torsion index gives some of the invariants of SQFTs suggested by topological modular forms,
and is expected to be zero for the cases that are relevant to actual heterotic string theories.

\end{titlepage}

\setcounter{tocdepth}{2}

\newpage

\tableofcontents

\section{Introduction and summary}
In the worldsheet formulation, general heterotic string theories are formulated
in terms of 2-dimensional $\cN=(0,1)$ supersymmetric quantum field theory (SQFT). For actual heterotic string theories,
SQFTs are required to be conformal and have specific left and right central charges $(c_L,c_R)$ so that 
it can be coupled to the worldsheet supergravity.
However, in this paper we consider general SQFTs whose worldsheet pure gravitational anomaly is specified by any integer $\nu \in \bZ$.
For the conformal case, $\nu$ is given by $\nu = 2(c_R - c_L)$.
However, we do not use nor assume conformal invariance at all throughout the paper.
One reason is that SQFTs themselves are interesting even without considering applications to heterotic strings.
Another reason is that such general SQFTs will be useful for the study of target space anomalies of actual heterotic strings.

This paper is devoted to a study of the following subjects.
\begin{enumerate}
\item The general structure of the target space topology/geometry of heterotic string theories.
\item Possible invariants of SQFTs other than the Witten index (elliptic genus) which forbids spontaneous supersymmetry breaking
in the infinite volume worldsheet.
\item Global anomalies of target space theories of general heterotic string theories. 
\end{enumerate}
We discuss them in Sec.\,\ref{sec:Bfield}, \ref{sec:torsion} and \ref{sec:anomaly}, respectively.
The second and third subjects are closely related. The first subject gives the basic setup for the study of the third one,
although technical details are not necessary.

The first subject about the topology/geometry of the target space of heterotic string theories is conceptually well understood~\cite{Witten:1999eg}, 
and our purpose is to clarify the general structure based on the recent understanding of anomalies.
We will consider arbitrary internal SQFT possibly with global symmetry $G$, 
and the target space is also general in the sense that it may not be orientable nor spin.
The principle for determining the target space structure is that we need to choose the structure so that there are no anomalies on the worldsheet
except for the pure gravitational anomaly $\nu \in \bZ\ \simeq (\IOspin)^4(\pt)$, where the notation $(\IOspin)^4(\pt)$ will be explained
later in the paper.
For example, the target space is often taken to be a spin manifold with the $B$-field satisfying $\d H = \lambda(R) - c(F)$,
where $H$ is the gauge invariant 3-form field strength of the $B$-field, $c(F) \sim \tr F^2$ is a characteristic class of the $G$-bundle
with the curvature 2-form $F$,
$\lambda(R)  = -\frac{1}{4(2\pi)^2} \tr R^2$ is one-half of the first Pontryagin class represented by the Riemann curvature 2-form $R$.
These conditions are imposed to avoid sigma model anomalies~\cite{Moore:1984dc,Moore:1984ws}.
Nonperturbative anomalies on the worldsheet have been studied in \cite{Witten:1985mj,Witten:1999eg},
and in particular the description of the $B$-field in \cite{Witten:1999eg} in terms of
the Dai-Freed theorem~\cite{Dai:1994kq}~\footnote{See \cite{Yonekura:2016wuc} for a physics explanation of the Dai-Freed theorem.} have led to the modern description of anomalies
of fermions in terms of a bulk theory in higher dimensions~\cite{Witten:2015aba,Witten:2016cio,Witten:2019bou}. This understanding is believed to be
valid for more general theories (see e.g. \cite{Freed:2014iua,Monnier:2019ytc}).
Along the lines of these developments, we give a systematic description of the $B$-field.
For example, the target space orientation and spin structure (or their generalization to non-orientable/non-spin cases) are naturally understood as part of the $B$-field. 
We will also demonstrate this understanding in the simpler case of supersymmetric quantum mechanics (SQM). 

Next let us discuss the second subject of the above list. 
The question about possible invariants of SQFTs other than the Witten index~\cite{Witten:1982df} has been raised.
It was conjectured~ \cite{Stolz:2004,Stolz:2011zj} that the ``space of $\cN=(0,1)$ SQFTs'' with the worldsheet pure gravitational anomaly $\nu \in \bZ$ is homotopy equivalent
to the $(-\nu)$-th space $\mathrm{TMF}_{-\nu}$ of a generalized cohomology theory known as topological modular forms, $\mathrm{TMF}$.
See \cite{Gaiotto:2018ypj,Gukov:2018iiq,Gaiotto:2019asa,Gaiotto:2019gef,Johnson-Freyd:2020itv,Tachikawa:2021mvw,Tachikawa:2021mby,Lin:2021bcp} for the physics literature discussing this conjecture. If the conjecture is correct,
it implies that there is new obstruction to spontaneous supersymmetry breaking
beyond the Witten index or the elliptic genus~\cite{Witten:1986bf}, no matter what continuous deformation we perform. 
This question has been investigated in \cite{Gaiotto:2019asa}, focusing
on the case of sigma models with the target space $S^3$ and any Wess-Zumino-Witten term. 
Then, a new invariant of SQFTs has been proposed by Gaiotto and Johnson-Freyd \cite{Gaiotto:2019gef}
that can explain the obstruction to supersymmetry breaking of the $S^3$ sigma models (and others) in the infinite volume.\footnote{
On the other hand, for a finite volume $\bR \times S^1$ where $\bR$ is time and $S^1$ is space, 
supersymmetry can be broken. Indeed, it happens in the $S^3$ sigma models studied in \cite{Gaiotto:2019asa}.
The point is that supersymmetry is broken by the energy scale of order the inverse radius of the space $S^1$,
and the breaking scale cannot be made parametrically larger than the inverse radius.
Thus the supersymmetry is restored in the infinite volume limit $S^1 \to R$.}
The invariant which we will discuss in the present paper is basically the same as the one in \cite{Gaiotto:2019gef},
but our definition and computational methods are different. The motivation of the present paper comes from the question of global anomalies of heterotic string theories,
and our definition is suited for that purpose and gives additional insight.  
The invariant is a version of the Witten index as will become clear later. 
But it is torsion, meaning that if we multiply it by some large enough integer
then it vanishes. Thus we call it the torsion index.

Finally, we discuss the third subject of the list.
Perturbative anomalies of the target space theories are well-understood. The Green-Schwarz mechanism~\cite{Green:1984sg}
is extended to arbitrary heterotic string theories with any internal theory that is not 
necessarily geometric~\cite{Schellekens:1986xh,Lerche:1987qk,Lerche:1988np}.
However, global anomalies are less understood. For example, it was investigated in the 10-dimensional $E_8 \times E_8$ 
heterotic string in \cite{Witten:1985bt}.
A question about general internal SCFTs has been recently raised in \cite{Enoki:2020wel},
and studied by using mathematical methods of algebraic topology in \cite{Tachikawa:2021mvw,Tachikawa:2021mby}.
In particular, under the aforementioned conjecture on $\mathrm{TMF}$ (as well as some additional assumptions),
it has been argued that the absence of global anomalies is translated to the fact that $\mathrm{TMF}^{21}(\pt)=0$~ \cite{Tachikawa:2021mby}.
In the present paper, we will give a different, more field-theoretical argument relating 
the target space global anomalies to $\mathrm{TMF}^{21}(\pt)$. 
For this purpose, we express the target space global anomalies by the torsion index of SQFTs whose worldsheet pure gravitational anomaly is
$\nu=-21$. Assuming the conjecture on $\mathrm{TMF}$, the fact $\mathrm{TMF}^{21}(\pt)=0$ implies that there is no such invariant and hence the torsion index
must be zero. We can also consider ``wrong'' heterotic string theories with different values of $\nu$ such that
the target space global anomalies do not vanish. As an example, we will understand the torsion index of the $S^3$ sigma models
as a global anomaly of a wrong 2-dimensional heterotic string theory.
The relevant global anomaly is the $\bZ_{24}$ anomaly found in \cite{Tachikawa:2021mvw},
and it is translated to elements of $\mathrm{TMF}^{-3}(\pt) = \bZ_{24}$.

\section{Structure of the {\it B}-field}\label{sec:Bfield}
In this section, we discuss general structure of the $B$-field in heterotic string theories.
The basic principle is the anomaly cancellation on the worldsheet,
and this requirement naturally gives the $B$-field.

\subsection{Review of anomalies}\label{sec:review}
Here we review the nonperturbative description of anomalies in terms of anomaly inflow from a higher dimensional bulk theory.
The following abstract discussion is more concretely understood 
in the case of fermions~\cite{Witten:1999eg,Witten:2015aba,Witten:2016cio,Witten:2019bou} 
(see also \cite{Witten:1985xe,Alvarez-Gaume:1984zst} for early work),
and it is believed to be valid for more general theories due to 
developments in condensed matter physics~\cite{Chen:2011pg,Kapustin:2014tfa,Kapustin:2014dxa}.

Suppose that we are interested in a $d$-dimensional theory with a symmetry group $H_d$.
Here $H_d$ includes Lorentz as well as internal symmetries. For example, in our later applications
to the $d=2$ worldsheet, we take $H_d = \Spin(d) \times G$, where $\Spin(d)$ is (the Euclidean version of) 
the $d$-dimensional Lorentz symmetry group, and $G$ is an internal symmetry group. 
More details on the sequence of groups $H_d~(d=1,2,\cdots)$ are explained in \cite{Freed:2016rqq}.

We turn on background fields for the symmetry $H_d$. More explicitly, the background fields
are a metric tensor and a connection of a $H_d$-bundle which is compatible with the metric.
We also assume that the theory can be coupled to a background sigma model field with target space $X$.
We only consider the case that $H_d$ acts trivially on $X$.
Whenever we consider a manifold $M_d$, we implicitly assume that it is equipped with these background fields
(i.e. an $H_d$ bundle $P \to M_d$, a metric, a connection compatible with the metric, and a sigma model field $f: M_d \to X$).
If we are not interested in sigma model background fields, we just take $X$ to be a point, $X=\pt$.

When the theory has an anomaly under the symmetry $H_d$ and/or the sigma model $X$,
the partition function (or un-normalized correlation functions) of the theory on $M_d$ is not defined as a complex number.
To remedy this, we introduce a $(d+1)$-manifold $N'_{d+1}$ such that
$M_d$ is realized as the boundary of $N'_{d+1}$, that is, $\partial N'_{d+1} =  M_d$.
Again, we always assume that the background fields on $M_d$ are extended to $N'_{d+1}$.
(The bundle on $N'_{d+1}$ is now an $H_{d+1}$-bundle since the dimension is increased. We often suppress $d$ and just write $H$
in the following.)
The case that we cannot take such an extension $N'_{d+1}$ will be discussed later.
We put some bulk action in $N'_{d+1}$ so that the total partition function of the bulk-boundary system is well-defined. 
For example, for fermion anomalies at the perturbative level, we may introduce Chern-Simons terms such as $\tr (A \wedge \d A + \frac{2}{3}A^3)$
in the case of $d=2$ where $A$ is a connection field. At the nonperturbative level, we use the Atiyah-Patodi-Singer (APS) $\eta$-invariant. 

The total partition function may depend on the choice of an extension $N'_{d+1}$.
If we take another extension $N''_d$, the difference of the partition functions between the two choices
is given as follows. We consider the closed manifold $N_{d+1} = N'_{d+1} \cup \overline{N''_{d+1}}$ that is obtained
by gluing $N'_{d+1} $ and the orientation reversal~\footnote{
More generally, even on nonorientable manifolds with some structure such as $\pin^\pm$ manifolds,
there is a generalization of the notion of orientation reversal. See \cite{Freed:2016rqq,Yonekura:2018ufj} for details. 
} 
$ \overline{N''_{d+1}}$ of $N''_{d+1}$ along the common boundary $M_d$.
Then the ratio between the two partition functions is given by the bulk partition function $Z(N_{d+1})$ on $N_{d+1}$.
We denote it as
\beq
Z(N_{d+1}) = \exp(2\pi \i\, h(N_{d+1}) ).
\eeq
The bulk partition function is a functional of the background fields, and for the purpose of anomaly inflow
we only need the case that $Z(N_{d+1}) \in \U(1)$ or in other words 
\beq
h(N_{d+1}) \in \bR/\bZ.
\eeq 
For example, in the case of $d=2$ fermions at the perturbative level, we may take 
$h(N_{3}) \sim \int_{N_{3}} \tr (A \wedge \d A + \frac{2}{3}A^3)$ up to coefficients which should be quantized
so that the values of $h(N_{d+1}) \in \bR/\bZ$ are gauge invariant. 
Nonperturbatively, $h(N_{d+1})$ is given by the APS $\eta$-invariant
of an appropriate Dirac operator in $N_{d+1}$, $h(N_{d+1}) = -\eta(N_{d+1})$. For Majorana fermions,
we instead have $h(N_{d+1}) = -\eta(N_{d+1})/2$.

In all known cases of unitary invertible anomalies, the function $h$ has the following property.
Associated to $h$, there exists a $(d+2)$-form $\omega$ of the form
\beq\label{eq:Apol}
\omega = \sum_{a} \omega_a c_a(F) 
\eeq
where $\{\omega_a\}$ are some closed differential forms $\d \omega_a=0$ on $X$,
$F$ is a formal 2-form variable that takes values in the Lie algebra of the symmetry group $H$,
and $\{c_a(F)\}$ are invariant polynomials of $F$.\footnote{More precisely, $\omega$
is a closed form on $X$ whose coefficients are invariant polynomials of the dual of the Lie algebra of $H$.
Then we can perform the Chern-Weil construction. More generally, $c_a(F)$ may change sign
under elements of $H$ which reverse orientation. See \cite{Yamashita:2021cao} for more details. 
}
We will take $F$ to be the curvature 2-form $F = \d A +A^2$
of the background connection of the $H$-bundle.
We will also pullback $\omega_a$ by the sigma model map $f : \textrm{(spacetime)} \to X$
to get differential forms on spacetime manifolds. Now, if $N_{d+1}$ is the boundary of a $(d+2)$-manifold $L_{d+2}$,
then $h$ satisfies
\beq\label{eq:compatibility}
h(\partial L_{d+2}) = \int_{L_{d+2}} \omega \mod \bZ, 
\eeq
where in the right hand side it is understood that $F$ is taken to be the curvature 2-form and we also pullback differential forms $\omega_a$ from $X$
to $L_{d+2}$ by the sigma model map.
For example, if we are interested in the trivial sigma model $X = \pt$, we take $\omega_a=1$.
In the example where $h(N_{3}) \sim \int_{N_{3}} \tr (A \wedge \d A + \frac{2}{3}A^3)$,
we have $\omega \sim \tr F^2$. 

Nonperturbatively, 
the APS index theorem states that the index of an appropriate Dirac operator $\cD_{d+2}$ in $(d+2)$-dimensions is given by
\beq
\index \cD_{d+2}(L_{d+2}) = \int_{L_{d+2}} \cI + \eta( \partial L_{d+2}),
\eeq
where $\cI$ is the characteristic polynomial of the gauge and Riemann curvature 2-forms that appears in the Atiyah-Singer index theorem
in $(d+2)$-dimensions. Therefore, if $h$ is given by the APS $\eta$-invariant as $h(N_{d+1}) = -\eta(N_{d+1})$, then we have $\omega = \cI$ since
the integer $\index \cD_{d+2}(L_{d+2}) \in \bZ$ drops out from the equation \eqref{eq:compatibility}.
For Majorana fermions we have $h(N_{d+1}) = -\eta(N_{d+1})/2$, but $\index \cD_{d+2}(L_{d+2}) \in 2\bZ$ and hence $\omega =\cI/2$ (see \cite{Witten:2019bou} for details). 
In general, we call $\omega$ the anomaly polynomial.

Let us summarize the properties of the bulk action. Let $\cC_{d+1}$ be the set of closed $(d+1)$-manifolds $N_{d+1}$ equipped with background fields.
This set is actually regarded as a monoid by the disjoint union of manifolds, and can also be made into an abelian group by  
introducing some equivalence relation. We denote this group by the same symbol $\cC_{d+1}$.
Then $h$ is a homomorphism
\beq
h : \cC_{d+1} \ni N_{d+1} \mapsto h(N_{d+1}) \in \bR/\bZ.
\eeq
Moreover, there exists $\omega$ associated to $h$ 
such that $h$ satisfies \eqref{eq:compatibility} for any $(d+2)$-manifold $L_{d+2}$ possibly with boundary.
The $(d+1)$-dimensional bulk theory is described by such a pair $(h,\omega)$.

The set of pairs $(h,\omega)$ satisfying the condition \eqref{eq:compatibility} forms an abelian group in the obvious way.
We denote this abelian group as 
\beq
\widehat{(\IO^H)}{}^{d+2}(X) =\{(h,\omega)\}
\eeq
This is the group whose element determine a $(d+1)$-dimensional bulk theory.

If $(h,\omega)=0$, the bulk theory is trivial and there is no anomaly in the boundary theory. However, even if it is nonzero,
we can still cancel anomalies if the following condition is satisfied. Suppose that there is a gauge invariant $(d+1)$-form $\alpha$
which is of the form \eqref{eq:Apol}, 
\beq
\alpha = \sum_a \alpha_a  c_a(F).
\eeq
However, we do not require it to be a closed form, so we may have $\d \alpha_a \neq 0$.
Chern-Simons differential forms like $\tr (A \wedge \d A + \frac{2}{3}A^3)$ are not gauge invariant and we do not consider them as $\alpha$.
Only manifestly gauge invariant $\alpha$ is allowed. 
We can consider $h_\alpha$ given by
\beq\label{eq:purediff}
h_\alpha (N_{d+1})  = \int_{N_{d+1} } \alpha. 
\eeq
A pair $(h_\alpha, \d \alpha)$ is an element of $\widehat{(\IO^H)}{}^{d+2}(X)$.
This type of bulk action does not contribute to the anomaly inflow because $\alpha$ itself is gauge invariant
and hence $\int_{N'_{d+1}} \alpha$ is gauge invariant even if $N'_{d+1}$ has a boundary. 
Therefore, we introduce an equivalence relation $\sim$ in the group $\widehat{(\IO^H)}{}^{d+2}(X) $ as
\beq
(h, \omega) \sim (h,\omega) + (h_\alpha, \d \alpha).
\eeq
These two pairs are related by a manifestly gauge invariant counterterm $\alpha$
and hence produce the same anomaly. We denote the group of equivalence classes as ${(\IO^H)}{}^{d+2}(X) $,
\beq
{(\IO^H)}{}^{d+2}(X) = \widehat{(\IO^H)}{}^{d+2}(X) / \sim.
\eeq
This group ${(\IO^H)}{}^{d+2}(X) $ classifies anomalies of $d$-dimensional theories with 
the symmetry group $H$ and the background sigma model with the target space $X$.
We denote the equivalence class of $(h,\omega) \in  \widehat{(\IO^H)}{}^{d+2}(X) $ as $[(h,\omega)] \in {(\IO^H)}{}^{d+2}(X)$.

Now, it is proved in \cite{Yamashita:2021cao} that ${(\IO^H)}{}^{\bullet}$ coincides
with a generalized cohomology theory known as the Anderson dual of the bordism theory $\Omega^H_{\bullet} $.\footnote{
The classification of invertible phases in terms of the Anderson dual of the bordism theory is conjectured in \cite{Freed:2016rqq}
and it motivated the work \cite{Yamashita:2021cao}. See also \cite{Lee:2020ojw} where the above interpretation of ${(\IO^H)}{}^{d+2}(X) $
is anticipated. 
}
The group ${(\IO^H)}{}^{d+2}(X)$ is the $(d+2)$-th cohomology group of the space $X$.
Here the bordism group $\Omega^H_d(X)$ is defined by introducing an equivalence relation in the set $\cC_d$
of $d$-manifolds $M_d$ equipped with an $H$-bundle and a sigma model map $f : M_d \to X$. 
Two manifolds $M_d$ and $M'_d$ are defined to be equivalent if there exists a $(d+1)$-manifold $N_{d+1}$
such that $\partial N_{d+1} = M_d \sqcup \overline{M'_d}$, where $\sqcup$ means the disjoint union, and the overline on $\overline{M'_d}$
means orientation reversal (or its generalization). We may denote the equivalence class as $[M_d]$ and define 
$\Omega^H_d(X) = \{ [M_d] \}$ to be the abelian group of equivalence classes.
The abelian group structure in $\Omega^H_d(X)$ is defined by the disjoint union as $[M_d]+[M'_d]= [M_d \sqcup M'_d]$,
the group unit is $[\varnothing]$, and the inverse of $[M_d]$ is $[\overline{M_d}]$.

The Anderson dual for generalized cohomology theories is defined in a very abstract way in mathematics and we do not explain it. 
However, the above definition of ${(\IO^H)}{}^{d+2}(X)$
gives a more intuitive understanding~\cite{Yamashita:2021cao,Lee:2020ojw}. In particular, we have a short exact sequence
\beq \label{eq:short}
0 \to \Ext( \Omega^H_{d+1}(X) , \bZ) \to {(\IO^H)}{}^{d+2}(X) \to \Hom(\Omega^H_{d+2}(X), \bZ) \to 0,
\eeq
where $\Ext(\bA,\bZ)$ for an abelian group $\bA$ is given by
\beq
\Ext( \bA, \bZ) =\Hom(\bA,\bR/\bZ)/\Hom(\bA,\bR).
\eeq
In other words, it is the torsion part of $\Hom(\bA, \bZ/\bR)$ if $\bA$ is finitely generated. (For example,
$\Hom(\bZ_k,\bR/\bZ) \simeq \bZ_k$, $\Hom(\bZ_k,\bR)\simeq 0$ and hence
$\Ext(\bZ_k, \bZ) \simeq \bZ_k$.). 
The meaning of this short exact sequence is as follows. Among the elements of $\widehat{(\IO^H)}{}^{d+2}(X)$,
there are elements such that the $(d+2)$-form $\omega$ is zero. The condition \eqref{eq:compatibility} for $\omega=0$ implies that $(h,0)$ 
gives an element of $\Hom(\Omega^H_{d+1}(X),\bR/\bZ)$,
and the equivalence class $[(h,0)]$ is an element of $ \Ext(\Omega^H_{d+1}(X) , \bZ) $.~\footnote{
The fact that invertible quantum field theories with $\omega=0$ 
are classified by $\Hom( \Omega^H_{d+1}(X) , \bR/\bZ)$ can be proved by starting from some axioms of
topological quantum field theory~\cite{Freed:2016rqq,Yonekura:2018ufj}. Non-torsion part is not yet proved from any axioms of quantum field theory,
so \cite{Yamashita:2021cao} started from the empirical fact expressed in \eqref{eq:compatibility}.  
}
This is the meaning of the term $ \Ext(\Omega^H_{d+1}(X) , \bZ) $ in \eqref{eq:short}. 
On the other hand,
given an arbitrary element $[(h,\omega)] \in {(\IO^H)}{}^{d+2}(X)$,
we can define an element of $\Hom(\Omega^H_{d+2}(X), \bZ)$ by $[L_{d+2}] \mapsto \int_{L_{d+2}} \omega$.
Notice that the condition \eqref{eq:compatibility}  for $\partial L_{d+2} = \varnothing$ implies that $\int_{L_{d+2}} \omega \in \bZ$.
This is the meaning of the term $\Hom(\Omega^H_{d+2}(X), \bZ)$ in \eqref{eq:short}.
One can check that the sequence \eqref{eq:short} is exact. 

We have mentioned that the partition function of a $d$-dimensional anomalous theory on $M_d$
is well-defined if we take $N_{d+1}$ such that $\partial N_{d+1} = M_d$.
If $[M_d]$ is a nontrivial element of $\Omega^H_d(X)$, it is not possible to find such $N_{d+1}$.
In this case, we need to fix the phase of the partition function ``by hand''. 
This is allowed by the following reason. If two manifolds are bordant, i.e. $[M_d] = [M'_d]$,
we can choose $N_{d+1}$ such that $\partial N_{d+1} = M_d \sqcup \overline{M'_d}$ and the partition function
for this configuration is already defined. From this consideration, one can see that 
the phase ambiguity is controlled by $\Hom(\Omega^H_d(X), \bR/\bZ)$.
More precisely, any two choices of the phases of the partition function is related by an element of $\Hom(\Omega^H_d(X), \bR/\bZ)$ as
\beq\label{eq:Theta}
\exp(2\pi \i\, \Theta(M_d)) , \qquad \Theta \in \Hom(\Omega^H_d(X), \bR/\bZ).
\eeq
Such terms are generalized $\theta$-angles of the background fields in $d$-dimensions,
and they are allowed by the principles of quantum field theory \cite{Kapustin:2014tfa,Kapustin:2014dxa,Freed:2016rqq,Yonekura:2018ufj}. 
Thus, we can fix the phase ambiguity in any way. The space of choices is a torsor over $ \Hom(\Omega^H_d(X), \bR/\bZ)$.

\subsection{The {\it B}-field and worldsheet anomalies}\label{sec:BW}
The understanding of anomalies as in the previous subsection has been already
recognized at least conceptually in the context of the $B$-field in heterotic string theories by Witten \cite{Witten:1999eg}.\footnote{
See also \cite{Distler:2009ri,Distler:2010an,Freed:2014iua} for a proposal for the $B$-field in the case of Type~II string theories.}
Our purpose is to describe it systematically for more general theories. 

Although our main interest is the $d=2$ worldsheet, we can also get some insight into the simpler case
of SQM ($d=1$), so we still take $d$ to be arbitrary.

We consider a theory $\cT$ with a global internal symmetry $G$.
The theory possibly depends on the spin structure, so the total symmetry group including the spacetime symmetry is $\Spin(d) \times G$.
This is the internal SQFT in the context of heterotic string theories. 
The $(d+1)$-dimensional bulk theory describing its anomaly is given by an element 
\beq
(h_\cT,\omega_\cT) \in \widehat{(\IO^{\spin \times G})}{}^{d+2}(\cP),
\eeq
where $\cP$ is a possible background sigma model coupled to $\cT$.
Intuitively, we regard it as the space of parameters (coupling constants) of $\cT$ which is promoted to a background sigma model.

Let us also consider $D$ scalar fields $\phi^I~(I=1,\cdots, D)$ and their superpartner fermions $\psi^I$
on which the symmetry group $\O(D)$ acts. We have in mind either SQM $d=1, \cN=1$ or SQFT $d=2, \cN=(0,1)$.
For a moment they are regarded as taking values in $\bR^D$.
The total symmetry group is $ \Spin(d) \times \O(D)$,
and its corresponding bulk theory is given by an element 
\beq
(h_\psi,\omega_\psi) \in \widehat{(\IO^{\spin \times \O(D)})}{}^{d+2}(\pt).
\eeq

Let $H_d = \Spin(d) \times G \times \O(D)$. Then we get $(h_\cT,\omega_\cT) +(h_\psi,\omega_\psi) \in   \widehat{(\IO^{H})}{}^{d+2}(\cP)$.
This sum may contain a pure gravitational anomaly which survives even if we take background fields other than the worldsheet metric to be trivial.
Let $(h_\textrm{pure}, \omega_\textrm{pure}) \in \widehat{(\IOspin)}{}^{d+2}(\pt)$ be the negative of the pure gravitational anomaly 
contained in $(h_\cT,\omega_\cT)+(h_\psi,\omega_\psi) $.
In the context of actual heterotic string theories, this is a contribution from the worldsheet gravitino of conformal supergravity.
In any case, we allow SQFTs to have pure gravitational anomalies,
and just use $(h_\textrm{pure}, \omega_\textrm{pure})$ to subtract pure gravitational anomalies.
We denote the sum of all contributions as
\beq\label{eq:beforesigma}
(h,\omega) = (h_\cT,\omega_\cT) +(h_\psi,\omega_\psi) +(h_\textrm{pure}, \omega_\textrm{pure}) \in \widehat{(\IO^{H})}{}^{d+2}(\cP).
\eeq
The pure gravitational anomaly of this sum is zero by definition. 

We want to couple the two theories $\cT$ and $(\phi,\psi)$ in the following way.
Consider a Riemannian $D$-manifold $X$ equipped with a $G$-bundle $P$ and its connection $A$, and also a map $g : X \to \cP$.
Now we take the target space of the scalar fields $\phi$ to be $X$ instead of $\bR^D$,
\beq
\phi : M_d \to X,
\eeq
where $M_d$ is the worldsheet (or worldline). 
The fermions $\psi$ are taken to be sections of the pullback of the tangent bundle, $\phi^*TX$.
We pullback the Levi-Civita connection $\Gamma$ on $X$ to get an $\O(D)$ connection $\phi^*\Gamma$ that is used for the covariant derivative of $\psi$.
Then $(\phi,\psi)$ is the supersymmetric sigma model with the target space $X$.
We also pullback the $G$-bundle $P$ and its connection $A$, and use the pullback $\phi^* A$ as a $G$-connection that is coupled to $\cT$.
We also take $\phi^* g = g \circ \phi : M_d \to \cP$ as a sigma model that can be coupled to $\cT$.
In this way, we get a sigma model with the target space $X$ that is coupled to the theory $\cT$.

We can consider the bulk theories $(\widetilde h_\cT, \widetilde\omega_\cT)$ and $(\widetilde h_\psi, \widetilde \omega_\psi)$ obtained in such a way that
the background fields are taken to be the pullback as described above. 
Before the path integral over $\phi$ is performed, we can regard $\phi$ as a background field.
Then these bulk theories are elements of $\widehat{(\IOspin)}{}^{d+2}(X)$,
\beq
(\widetilde h_\cT, \widetilde \omega_\cT), \quad (\widetilde h_\psi, \widetilde \omega_\psi)~ \in~ \widehat{(\IOspin)}{}^{d+2}(X).
\eeq
We also denote $(h_\textrm{pure}, \omega_\textrm{pure})$ as
$(\widetilde h_\textrm{pure}, \widetilde \omega_\textrm{pure}) $ when it is regarded as an element of $ \widehat{(\IOspin)}{}^{d+2}(X)$.
Since $\phi$ is still a background field, two theories $\cT$ and $\psi$ are decoupled
and hence we can just add these contributions to get an element
\beq
(\widetilde h,\widetilde \omega) = 
(\widetilde h_\cT, \widetilde \omega_\cT)+(\widetilde h_\psi, \widetilde \omega_\psi)+ (\widetilde h_\textrm{pure}, \widetilde \omega_\textrm{pure}) 
\in \widehat{(\IOspin)}{}^{d+2}(X).
\eeq
By definition of $(h_\textrm{pure}, \omega_\textrm{pure})$, the pure gravitational anomaly is cancelled.

Now we can describe the precise structure of the target space $X$ and the $B$-field on it. 
For the path integral over $\phi$ to be consistent, anomalies must be absent except possibly for a pure gravitational anomaly.
However, we have added $ (\widetilde h_\textrm{pure}, \widetilde \omega_\textrm{pure}) $ so that there is no pure gravitational anomaly,
and hence all anomalies must vanish.\footnote{Throughout the paper we implicitly assume that anomaly-free fields can be path-integrated.}

At the topological level, this means that the equivalence class 
\beq
[(\widetilde h,\widetilde \omega) ] \in  {(\IOspin)}{}^{d+2}(X)
\eeq
must vanish. This topological condition may be rephrased in the following way.
We have defined \eqref{eq:beforesigma} before going to the sigma model with the target space $X$.
The topological class of this element is 
$[(h,\omega)] \in {(\IO^{H})}{}^{d+2}(\cP)$.
Let $BG$ and $B\O_D$ be the classifying space for $G$ and $\O(D)$, respectively.\footnote{
A classifying space $BG$ of a group $G$ is a topological space with the following properties. 
Let $[M, BG]$ be the set of homotopy classes of maps $f : M \to BG$ from a manifold $M$ to $BG$.
Then, there is one to one correspondence between elements of $[M,BG]$ 
and isomorphism classes of $G$-bundles on $M$. The correspondence is given as follows.
There is a $G$-bundle $G \to P_\text{univ} \to BG$, called the universal bundle.
Then the $G$-bundle $P$ on $M$ corresponding to $f \in [M,BG]$ is given by the pullback $P = f^* P_\text{univ}$. 
}
Then we have 
\beq
{(\IO^{H})}{}^{d+2}(\cP) = {(\IOspin)}{}^{d+2}(BG \times B\O_D \times \cP)
\eeq
since a manifold $M$ equipped with a $G$-bundle $P$ is topologically equivalent to the same manifold $M$ 
equipped with a map $f : M \to BG $ by the definition of the classifying space. 
Thus we get
\beq
[(h,\omega)]  \in {(\IOspin)}{}^{d+2}(BG \times B\O_D \times \cP).
\eeq
This is a kind of a characteristic class in $BG \times B\O_D \times \cP$, but the relevant cohomology theory here is not the ordinary cohomology $H^{\bullet}$
but a generalized cohomology ${(\IOspin)}{}^\bullet$. 
By the tangent bundle $TX$, the $G$-bundle $P$, and the map $g: X \to \cP$,
we have a classifying map 
\beq
f : X \to BG \times B\O_D \times \cP.
\eeq
This map can be used to pullback $[(h,\omega)]  $ to get a characteristic class in $X$ as
\beq
f^*[(h,\omega)]  \in {(\IOspin)}{}^{d+2}(X).
\eeq
This coincides with $[(\widetilde h,\widetilde \omega)]$,\footnote{${(\IOspin)}{}^{\bullet}$ is a functor
and hence it behaves naturally with respect to maps of the target spaces like $X \to X'$. 
Then we  may consider a manifold approximation to $BG \times B\O_D$ and consider a sigma model with that target space,
and pull back it to $X$.}
and the topological condition on $X$ is that this characteristic class must vanish.
We discuss when this is satisfied later.

Assume that the topological condition 
\beq \label{eq:topocondition}
[(\widetilde h,\widetilde \omega) ] =f^*[(h,\omega)] =0 \in   {(\IOspin)}{}^{d+2}(X)
\eeq
 is satisfied.
Next we need to add an appropriate counterterm so that the geometric quantity 
$(\widetilde h,\widetilde \omega) \in  \widehat{(\IOspin)}{}^{d+2}(X)$ is really cancelled. 
By the definition of $ {(\IOspin)}{}^{d+2}(X)$ in terms of $ \widehat{(\IOspin)}{}^{d+2}(X)$ discussed in Sec.~\ref{sec:review},
the fact that the topological class $[(\widetilde h,\widetilde \omega) ] $ is zero implies that
there exists a $(d+1)$-form $H$ such that 
\beq \label{eq:diffcondition}
(\widetilde h,\widetilde \omega) = -(h_H, \d H)
\eeq
where $h_H$ is given by
\beq
h_H(N_{d+1}) = \int_{N_{d+1}}H.
\eeq 
In dimensions $d+1 \leq 3$, we cannot use the worldsheet metric for $H$ because the lowest degree form constructed from the metric is 
a 4-form $\tr R^2$. Thus, $H$ is purely a differential form on $X$.
Now we add the action $h_H$ (multiplied by $2\pi \i $) to the bulk theory.
Then the anomaly is completely cancelled since $\widetilde h + h_H=0$.
In general, there is no canonical choice for $H$ and hence we need to choose $H$ as part of the data of the definition of the theory.
 
In the context of heterotic string theories, $H$ is the 3-form field strength of the $B$-field.
Notice that if some $H_0$ satisfies the condition $(\widetilde h,\widetilde \omega) = -(h_{H_0}, \d {H_0})$,
and if $H_1$ is a closed differential form representative of an integer cohomology class in $H^{d+1}(X,\bZ)$,
then $H=H_0 + H_1$ also satisfies the anomaly cancellation condition.
For example, for any 2-form $B'$, we can add $\d B'$ to the bulk $N'_{d+1}$.
When $N'_{d+1}$ has a boundary $\partial N'_{d+1} = M_d$, we get
\beq
\int_{N'_{d+1}} \d B' = \int_{M_d} B'.
\eeq
This is the usual term of the $B$-field in the worldsheet action at the perturbative level. 
Nonperturbatively, we cannot always write $H$ as a total derivative and hence we need
the extension of the worldsheet $M_d$ to a higher dimensional manifold $N_{d+1}$. 
The anomaly cancellation guarantees that the result does not depend on the choice of $N_{d+1}$.

We need to choose more data to define the theory.
When $[M_d] \in \Omega^\textrm{spin}_{d}(X)$ is nontrivial,
we choose the phase of the partition function. The space of such choices is a torsor
over $\Hom(\Omega^\textrm{spin}_{d}(X), \bR/\bZ)$ as we discussed around \eqref{eq:Theta}.
The group $\Omega^\textrm{spin}_{1}(X)$ is given by
\beq \label{eq:omega1}
\Omega^\textrm{spin}_{1}(X) \simeq H_0(X,\bZ_2) \oplus H_1(X,\bZ) .
\eeq
For $\Omega^\textrm{spin}_{2}(X)$, we have a short exact sequence
\beq \label{eq:omega2}
0 \to  H_0(X,\bZ_2)  \oplus H_1(X, \bZ_2) \to \Omega^\textrm{spin}_{2}(X)  \to H_2(X, \bZ) \to 0
\eeq
where $H_0(X,\bZ_2) $ is a direct summand.
These facts can be seen as follows. 
(We only sketch the argument. See \cite{Garcia-Etxebarria:2018ajm} for an explanation of how to use Atiyah-Hirzebruch spectral sequences
for the computation of bordism groups.)
Without loss of generality we may assume that $X$ is connected.
Let $\pt$ be an arbitrary point on $X$. Then, by using the maps $\pt \to X \to \pt$, we get
$ \Omega^\textrm{spin}_{d}(\pt) \to \Omega^\textrm{spin}_{d}(X) \to  \Omega^\textrm{spin}_{d}(\pt)$
and hence $\Omega^\textrm{spin}_{d}(X)$
contains $\Omega^\textrm{spin}_{d}(\pt)$ as a direct summand, 
\beq\label{eq:directsummand}
\Omega^\textrm{spin}_{d}(X) \simeq \Omega^\textrm{spin}_{d}(\pt) \oplus \widetilde {\Omega^\textrm{spin}_{d}}(X),
\eeq
where $ \widetilde {\Omega^\textrm{spin}_{\bullet}}$ is the reduced group.
Then we use Atiyah-Hirzebruch spectral sequence for $ \widetilde {\Omega^\textrm{spin}_{\bullet}}(X)$
whose $E^2$-term is $E^2_{p,q}= \widetilde{H_p}(X, \Omega^\textrm{spin}_{q}(\pt) )$.
It is known~\cite{Teichner:1993aij} that the differentials $d_2 : E^2_{p,q} \to E^2_{p-2,q+1}$ for $q=0,1$ are duals of 
the Steenrod square $Sq^2$ (after $\bZ_2$ reduction for the case $q=0$). By using 
\beq
\Omega^\textrm{spin}_{0}(\pt) \simeq \bZ, \quad \Omega^\textrm{spin}_{1}(\pt) \simeq \bZ_2, \quad
\Omega^\textrm{spin}_{2}(\pt) \simeq \bZ_2, \quad  \Omega^\textrm{spin}_{3}(\pt)=0, \quad  \Omega^\textrm{spin}_{4}(\pt)=\bZ
\eeq
we see that the differential $d_2$ is zero in the range of dimensions of our interest.
By dimensional reasons all other differentials $d_r : E^2_{p,q} \to E^2_{p-r,q+r-1} ~(r \geq 3)$ are also zero, 
so the spectral sequence converges already at the $E_2$-term. 
Notice also that the direct summand discussed above is $\Omega^\textrm{spin}_{d}(\pt) = H_0(X,\Omega^\textrm{spin}_{d}(\pt))$.
Thus we get the desired results for $\Omega^\textrm{spin}_{1}(X) $ and $\Omega^\textrm{spin}_{2}(X) $.

In the context of heterotic string theories, the meaning of the additional choice of the phase is
as follows. In the above discussion, we have only chosen the field strength 3-form $H$. 
However, $H$ does not completely determine the $B$-field. For instance, suppose that $H=0$. This means that the $B$-field is flat.
Thus we need to choose such a flat $2$-form. This is the information contained in $\Hom(H_2(X, \bZ), \bR/\bZ)$.
There is also other information. One (but not the only) way to satisfy the topological condition \eqref{eq:topocondition}
is to require that the manifold $X$ admits an orientation and a spin structure (as well as another condition about
the characteristic class at degree 4) as we discuss later. However,
we have not yet chosen any explicit orientation nor spin structure on $X$. The choice of orientation and spin structure is 
related to the topological terms $\Hom(H_0(X, \bZ_2), \bR/\bZ) \simeq H^0(X,\bZ_2)$ and $\Hom(H_1(X, \bZ_2), \bR/\bZ) \simeq H^1(X,\bZ_2)$, 
respectively. We will discuss a little more detail later for SQM in Sec.~\ref{sec:SQM}.
However, we remark that the target space $X$ need not be orientable nor spin in general,
depending on the internal theory $\cT$, its symmetry $G$ and anomalies.

\subsection{Supersymmetric quantum mechanics} \label{sec:SQM}
To illustrate the formulation of the previous subsection, 
let us consider the case of a simple SQM (i.e. $d=1, \cN=1$) in more detail. We take the internal theory $\cT$ to be trivial
and hence $G=1$ and $\cP = \pt$. We consider a target space $X$ whose dimension is $D = \dim X$. 
Anomalies of quantum mechanics are discussed e.g. in \cite{Elitzur:1985xj,Freed:2014iua,Witten:2019bou}. 
We will naturally obtain a $\spin^c$ connection on $X$ from the worldline perspective. 

The Lagrangian of the fermions  $\psi^I~(I=1,\cdots, D)$ with the $\O(D)$ symmetry is given by
\beq
\cL_\psi = \frac{\i}{2} \psi^I \frac{\d}{\d t}\psi^I.
\eeq
The canonical anticommutation relation after quantization is $\{\psi^I, \psi^J\}= \delta^{IJ}$.
These fermions have anomalies $[(h,\omega)] \in (\IOspin)^3(BO_D)$.
We want to impose the topological condition \eqref{eq:topocondition} on the characteristic class $f^*[(h,\omega)] $, where $f$ is the classifying map
$f : X \to BO_D$ determined by the tangent bundle $TX$.

Because $\Omega^\textrm{spin}_3(BO_D)$ is torsion and hence $\Hom(\Omega^\textrm{spin}_3(BO_D),\bZ)=0$, the short exact sequence
\eqref{eq:short} gives
\beq
(\IOspin)^3(BO_D) \simeq \Ext( \Omega^\textrm{spin}_{2}( BO_D ) , \bZ).
\eeq
Notice also that $\Omega^\textrm{spin}_{2} (BO_D)$ satisfies \eqref{eq:omega2} with the replacement $X \to BO_D$.

Without loss of generality, we assume that $X$ is connected. Then, $H_0(X,\bZ_2) \simeq \bZ_2$
and it actually comes from $ \Omega^\textrm{spin}_{2}(\pt) \simeq \bZ_2$. 
The anomaly associated to $\Ext(\Omega^\textrm{spin}_{2}(\pt), \bZ)  \simeq \bZ_2$
is just a pure gravitational anomaly in $d=1$. If the number of fermions $D$ is odd, the fermion parity $(-1)^F$ is anomalous.
Throughout the paper, we allow pure gravitational anomalies.

Next let us consider the anomaly associated to
\beq
 \Ext( H_1(BO_D, \bZ_2) , \bZ) \simeq H^1(BO_D, \bZ_2) \simeq \bZ_2,
\eeq
where the generator of $H^1(BO_D, \bZ_2) \simeq \bZ_2$ is the first Stiefel-Whitney class $w_1$.
This anomaly can be explicitly seen as follows. We consider the path integral of fermions $\psi^I$
on $S^1$ with the anti-periodic (bounding, NS) spin structure.
We put an $\O(D)$ holonomy $g$ around $S^1$ such that $\det g = -1$.
For instance, we can just take $g = \diag(-1,+1,\cdots,+1)$.
Then we have odd numbers of fermion zero modes, and hence the path integral measure is not invariant under 
the fermion parity $(-1)^F$. We conclude that our fermions $\psi^I$ have an anomaly.

The pullback $f^* w_1 \in H^1(X, \bZ_2) $ is the first Stiefel-Whitney class of the tangent bundle $TX$.
Thus, to avoid the above anomaly associated to $w_1$, we impose the condition that $X$ is orientable.
We remark that this conclusion is only because we have taken
the internal theory $\cT$ to be trivial. For example, if we take $\cT$ to be a single Majorana fermion (or fermi multiplet in $\cN=1$ supersymetry)
which transforms under $g \in \O(D)$ as $\det g$, then we can cancel the anomaly of $\psi^I$.
Then $X$ need not be orientable.

Assume that $X$ is orientable. As we discuss later, a choice of a topological term will give an explicit orientation on $X$, so let us assume that $X$ is oriented. 
Then, we can use $\SO(D)$ instead of $\O(D)$. Let $B\SO_D$ be a classifying space for $\SO(D)$.
Now we consider the anomaly associated to
\beq
\Ext( H_2(B\SO_D, \bZ) , \bZ) \simeq H^2(B\SO_D, \bR/\bZ)/ H^2(B\SO_D, \bR) \simeq \bZ_2.
\eeq
This $\bZ_2$ is generated by the second Stiefel-Whitney class $w_2$ regarded as an element of the
group $H^2(B\SO_D, \bR/\bZ)/H^2(B\SO_D, \bR)$. The fermions $\psi^I$ have the anomaly corresponding to this nontrivial element. 
A simple way to see this anomaly is to quantize the fermions $\psi^I$ and construct the Hilbert space.
Then the Hilbert space is in a spinor representation of $\Spin(D)$, and $\SO(D)$ acts as a projective representation.
Thus $\SO(D)$ is anomalous. Another way to see the anomaly is to repeat the original argument of Witten for 
the $d=4$ $\SU(2)$ anomaly~\cite{Witten:1982fp} in the current case of $d=1$ and $\SO(D)$, by using $\pi_1(\SO(D)) \simeq \bZ_2$. In any case,
we find that the fermions $\psi^I$ have the nontrivial anomaly.  

The corresponding sigma model anomaly is given by the pullback $f^* w_2$. However,
it is important that this pullback is regarded as an element of 
\beq
\Ext(H_2(X,\bZ), \bZ) \simeq  H^2(X, \bR/\bZ)/ H^2(X, \bR)
\eeq
rather than $H^2(X, \bZ_2)$. In more detail, the anomaly can be cancelled under the following condition.
The $2$-dimensional bulk theory is given by $\exp ( \pi \i \int f^* w_2)$,
and it is cancelled if and only if there exists a differential 2-form $F$ on $X$ such that
\beq\label{eq:spincF}
\exp ( \pi \i \int_{N_2} f^* w_2) = \exp ( -2\pi \i \int_{N_2} F)
\eeq
for closed manifolds $N_2$.
In this case $f^* w_2$, regarded as an element of $H^2(X, \bR/\bZ)$, actually comes from (the image of) $H^2(X, \bR)$,
and $F$ is a differential form representing it. This is exactly the condition for the existence of
a $\spin^c$ connection on $X$, and $F$ is a curvature 2-form.
By adding the counterterm $(h_{F}, 0)$ to the bulk theory in the notation of \eqref{eq:purediff},
the anomaly is cancelled. (Notice that $\d F=0$ because of \eqref{eq:spincF}.)
A choice of $F$ satisfying the above condition is part of the data of the theory.

We have seen the anomaly cancellation condition.
Now let us discuss topological terms 
\beq
\Hom(\Omega^\textrm{spin}_{1}(X), \bR/\bZ) \simeq H^0(X,\bZ_2) \oplus H^1(X, \bR/\bZ).
\eeq
where we have used \eqref{eq:omega1}.

For connected $X$, we have $ H^0(X,\bZ_2) \simeq \bZ_2$, which actually comes from $ \Hom(\Omega^\textrm{spin}_{1}(\pt), \bR/\bZ)$.
In the path integral, the effect of the nontrivial element of $ \Hom(\Omega^\textrm{spin}_{1}(\pt), \bR/\bZ) \simeq \bZ_2$ is that 
it assigns $(-1)$ to $S^1$ with the periodic (non-bounding, Ramond) spin structure. 
In other words, it assigns additional sign to $\tr (-1)^F$. Thus, the fermion parity $(-1)^F$ of the states is flipped. 
This is related to the orientation of $X$ by the following reason. As we mentioned above,
the quantization of $\psi^I$ gives a spinor representation of the Lie algebra $\so(D)$ of $\SO(D)$. 
(This is actually a representation of $\Spin^c(D) = (\Spin(D) \times \U(1))/\bZ_2$.)
States with $(-1)^F=+1$ are said to have positive chirality, while those with $(-1)^F=-1$ are said to have negative chirality. 
If we add $ \Hom(\Omega^\textrm{spin}_{1}(\pt), \bR/\bZ)$, positive chirality and negative chirality are exchanged. 
This means that the orientation of $X$ is changed. Notice that at the beginning
there is no canonical way to specify which is positive chirality and which is negative chirality.
Thus the choice of orientation is a torsor over $H^0(X,\bZ_2)$.

Next consider $ H^1(X, \bR/\bZ)$. If we add a topological term $A_{\rm flat} \in H^1(X, \bR/\bZ)$
then in the path integral on $S^1$, we get an additional phase factor $\exp(2\pi \i \int A_{\rm flat})$.
This is a flat part of the $\spin^c$ connection. The non-flat part is specified by the curvature 2-form $F$ introduced above,
but the curvature 2-form does not completely determine a $\spin^c$ connection. It is specified by the phase choice which is
a torsor over $H^1(X, \bR/\bZ)$. In this way, we get a complete $\spin^c$ connection on the target space $X$. 
Indeed, the path integral of the fermions under fixed bosonic fields $\phi^I$ gives the Wilson loops
of the $\spin^c$ connection.

So far we have assumed that the Lorentz symmetry of the worldline is $\Spin(d)$.
Let us briefly mention what happens if we include time-reversal symmetry $\sT$ with 
\beq
\sT^2=1, \qquad \sT \psi^I \sT^{-1}=\psi^I \quad (I=1,\cdots,D).
\eeq
In Euclidean spacetime, it corresponds to the symmetry group $\Pin^-(d)$ (see \cite{Kapustin:2014dxa,Witten:2015aba} for details.)
We have
\beq
\Omega^{\pin^-}_{0}(\pt) \simeq \bZ_2, \quad \Omega^{\pin^-}_{1}(\pt) \simeq \bZ_2, \quad
\Omega^{\pin^-}_{2}(\pt) \simeq \bZ_8, \quad  \Omega^{\pin^-}_{3}(\pt)=0.
\eeq
Let us only mention the main differences from the case of $\Spin(d)$.

The pure gravitational anomaly is $\Ext(\Omega^{\pin^-}_{2}(\pt), \bZ) \simeq \bZ_8$ rather than 
$\Ext(\Omega^{\spin}_{2}(\pt), \bZ) \simeq \bZ_2$. 
The time reversal 
symmetry $\sT$ gives a real structure to the Clifford algebra $\{\psi^I, \psi^J\} = \delta^{IJ}$, and $\bZ_8$
is the famous mod 8 periodicity of the real representations of the Clifford algebra (see e.g. \cite{Weinberg:2000cr} for a review). 

The anomaly associated to $f^*w_2$ should be now regarded as an element of $H^2(X,\bZ_2)$ rather than $H^2(X,\bR/\bZ)/H^2(X,\bR)$
because of the difference between $\Omega^{\pin^-}_{0}(\pt) \simeq \bZ_2$ and $\Omega^{\spin}_{0}(\pt) \simeq \bZ$.
Thus, $f^* w_2$ must be zero and hence $X$ must admit a spin structure rather than $\spin^c$.

One of the topological terms is classified by $H^1(X, \bZ_2)$ rather than $H^1(X, \bR/\bZ)$ because of the difference between 
$\Omega^{\pin^-}_{0}(\pt) \simeq \bZ_2$ and $\Omega^{\spin}_{0}(\pt) \simeq \bZ$. This is related to the spin structure on the target space $X$.

\subsection{Heterotic string theories}
Let us return to the case of $d=2, \cN=(0,1)$ theories, with an arbitrary internal theory $\cT$.
The fermions $\psi^I$ are right-moving chiral fermions on the worldsheet.
In general, the topological condition \eqref{eq:topocondition} is complicated.
Thus we only discuss a few points. For simplicity we only consider the case $\cP = \pt$.

Let $\lambda(R) = -\frac{1}{4(2\pi)^2} \tr R^2$ be one-half of the first Pontryagin class represented by the Riemann curvature 2-form $R$
of the target space $X$.
Let $c(F)$ be the anomaly polynomial of the theory $\cT$ under the symmetry $G$, where $F$ is the curvature 2-form of the $G$-bundle on $X$.
(Roughly $c(F) \sim \tr F^2$.)
The differential form part of the anomaly cancellation condition \eqref{eq:diffcondition} is given by
\beq \label{eq:dffRFcondition}
\d H = \lambda(R)  - c(F). 
\eeq
Here $\lambda(R)$ is the contribution from the anomaly of the fermions $\psi^I$ under $\O(D)$, 
and $c(F)$ is that from the anomaly of the theory $\cT$ under $G$. 
This is the famous condition in heterotic string theories at the differential form level.
However, there is also conditions from the part which cannot be expressed by differential forms.

As a simple case, suppose that $G$ is connected and simply connected, $\pi_0(G)=0, ~\pi_1(G)=0$.
In this situation, we can solve the condition \eqref{eq:topocondition} as follows.

First, $X$ must admit orientation and spin structure. This can be seen
as in the case of SQM in Sec.~\ref{sec:SQM} with the time reversal symmetry.
In fact, we can reduce a $d=2$ theory to a $d=1$ theory by compactification on $S^1$ that has the periodic (non-bounding, Ramond)
spin structure. The group $G$ does not have any anomaly in $d=1$ when it is connected and simply connected,
and hence there is no contribution from the theory $\cT$ in $d=1$ (except possibly for pure gravitational anomalies). 
Moreover, the $\mathsf{CPT}$ symmetry in $d=2$ reduces to the time reversal symmetry $\sT$ in $d=1$ as
$\mathsf{CPT}_{d=2} = \sT_{d=1}$. Therefore, the results of the previous subsection apply directly. 
(One can also directly work in $d=2$.)

Thus let us assume that $X$ admits orientation and spin structure.
As mentioned in Sec.~\ref{sec:SQM}, $X$ is equipped with an orientation and a spin structure
by specifying the topological terms classified by $H^0(X,\bZ_2) \oplus H^1(X,\bZ_2)$.
We can now take the symmetry group of $\psi^I$ to be $\Spin(D)$, which is connected and simply connected (assuming $D \geq 3$).
Then, under the condition $\pi_0(G)=\pi_1(G)=0$ we have (from Atiyah-Hirzebruch spectral sequence) 
\beq
{(\IOspin)^4}(BG \times B\Spin_D) \simeq {(\IOspin)^4}(\pt) \oplus H^4(BG \times B\Spin_D),
\eeq
where $B\Spin_D$ is the classifying space for $\Spin(D)$.
Here ${(\IOspin)^4}(\pt) \simeq \bZ $ corresponds to pure gravitational anomalies and 
it is irrelevant for the condition \eqref{eq:topocondition} (see Sec.~\ref{sec:BW}).
Then the class $[(h,\omega)] \in (\IOspin)^4(BG \times B\Spin_D) $
that appears in the condition \eqref{eq:topocondition} is
actually given by $\lambda - c$, where $\lambda \in H^4(B\Spin_D)$ is the generator of $ H^4(B\Spin_D) \simeq \bZ$
given by one-half of the first Pontryagin class $p_1$, and $c \in H^4(BG,\bZ) \simeq \bZ$ is a characteristic class (not necessarily a generator)
whose differential form representative is $c(F)$. Thus the topological condition is
\beq\label{eq:ordinarycoho}
0=f^*(\lambda-c) \in H^4(X,\bZ).
\eeq
Notice that \eqref{eq:dffRFcondition} ensures this condition at the level of de~Rham cohomology,
but this topological condition must be satisfied at the level of integral cohomology.
The condition at the integral cohomology level was noticed in \cite{Witten:1985mj}.
The necessity of this condition can also be seen from the perspective of the target space $p$-form gauge theories as
discussed in \cite{Hsieh:2020jpj}.

We remark that \eqref{eq:dffRFcondition} does not give the complete condition for the differential form $H$.
The anomaly cancellation condition \eqref{eq:diffcondition} requires $H$ to satisfy
\beq
\int_{N_3} H \in - \widetilde h(N_3) + \bZ.
\eeq
This is a generalized version of the Dirac charge quantization condition for $H$.
If there were no anomaly, the Dirac quantization is simply given by $\int H \in \bZ$.
However, the anomaly $\widetilde h$ modifies the quantization condition. 

More generally, we can consider $G$ that is neither connected nor simply connected.
As an illustration, suppose that the theory $\cT$ consists of $D$ left moving fermions (or fermi multiplets) with the symmetry $G = \O(D)$.
Then we can cancel the anomalies between the left and right moving fermions by taking the $G$ bundle to the same
as the $\O(D)$ bundle associated to the tangent bundle of $X$. Then $X$ can be an arbitrary manifold.
Another example is to take $\cT$ to be the sigma model with the target space $S^1$ on which $\bZ_2$ symmetry acts by orientation reversal.
Then we can take $X$ so that the total theory is a sigma model whose target space is an $S^1$-bundle over $X$.
If $X$ is a $\pin^-$ manifold, we can introduce a spin structure to the total space.
Also by imposing the condition \eqref{eq:ordinarycoho} on the total space,
we get an anomaly-free theory. Considering $\pin^-$ manifolds can be interesting
because it gives more anomalies and constraints. See e.g. \cite{Montero:2020icj} for such constraints.

\section{The torsion index of SQFTs}\label{sec:torsion}
In this section, we introduce the notion of torsion index of 2-dimensional $\cN=(0,1)$ SQFTs.
As mentioned in the introduction, it is basically the same as the invariant proposed 
by Gaiotto and Johnson-Freyd \cite{Gaiotto:2019gef}.
However, we define it in a way that is more suitable for our later applications 
to the target space global anomalies of heterotic string theories.
Our definition also suggests different computational methods and we will discuss examples in Sec.~\ref{sec:anomaly}.

We will avoid concepts that are valid only for CFT. In particular, instead of Virasoro algebras, we just use
\beq
H_L = \frac{1}{2} (H + P), \qquad H_R = \frac{1}{2} (H - P).
\eeq
Here $H$ and $P$ are the Hamiltonian and the momentum operators on $\bR \times S^1$,
where $\bR$ is time, and $S^1$ is space with circumference $2\pi$ and the periodic (non-bounding, Ramond) spin structure
unless otherwise stated.
The supercharge $ Q$ of $\cN=(0,1)$ supersymmetry is such that 
\beq
Q^2= H_R.
\eeq

\subsection{Noncompact SQFTs and their index}\label{sec:noncompact}
Before discussing the torsion index, we need some preparation.
We consider ``mildly noncompact'' SQFTs as discussed in \cite{Gaiotto:2019asa,Gaiotto:2019gef,Johnson-Freyd:2020itv}.
We do not attempt to give a rigorous definition of this concept, but it is easy to understand intuitively. 

First, compact SQFTs are those for which the energy spectrum is discrete and sparse enough so that the
trace of the operator $q^{H_L} \bar{q}^{H_R}~~(q =e^{2\pi \i \tau})$ 
is well-defined for any complex modulus $\tau$ with $\Im \tau >0$.

\begin{figure}
\centering
\includegraphics[width=0.7\textwidth]{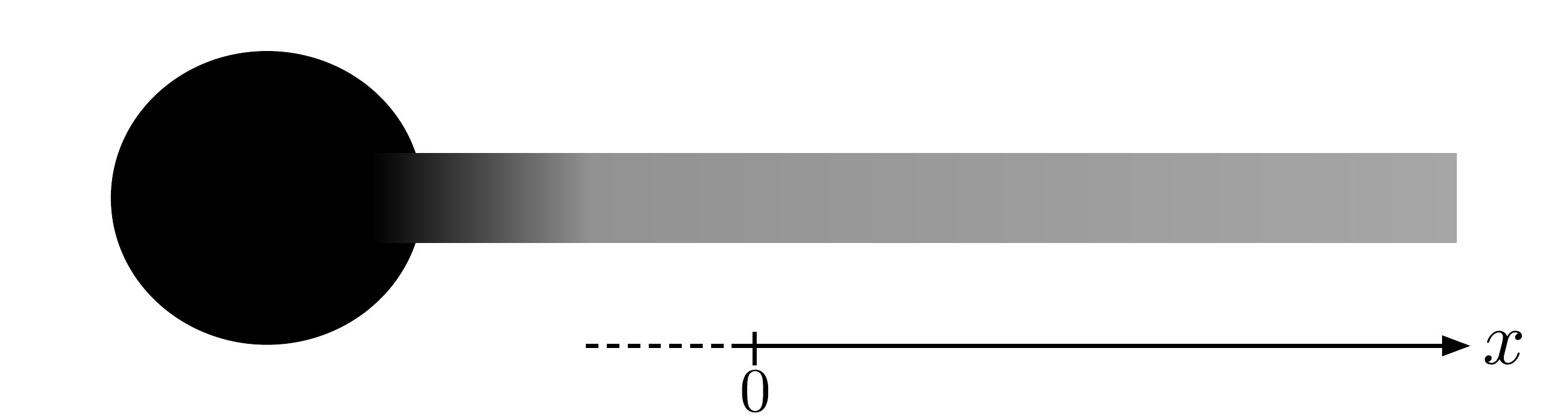}
\caption{Mildly noncompact SQFT. The black region is compact and it is connected
to the grey region in which the theory is of the form $\sigma(\bR_{\geq}) \otimes \cY$, where $\sigma(\bR_{\geq})$
is the sigma model with the target space $\bR_{\geq}=\{ x \in \bR~|~ x \geq 0\}$ (i.e. semi-infinite real line) and $\cY$ is a compact SQFT.
 \label{fig:nonC}}
\end{figure}

A mildly noncompact SQFT $\cZ$ is an SQFT in which the compactness is violated in a controlled way, as schematically shown in Fig.~\ref{fig:nonC}.
In the figure, the black region is compact. It is connected to the grey semi-infinite cylindrical region.
The theory in the grey region is of the form $\sigma(\bR_{\geq}) \otimes \cY$,
where $\sigma(\bR_{\geq}) $ is the sigma model with the target space $\bR_{\geq} =\{ x \in \bR~|~ x \geq 0\}$,
and $\cY$ is a compact SQFT. We call $\cY$ the boundary theory of $\cZ$.

A class of examples of mildly noncompact SQFTs is given by sigma models with a target space $X$
which we denote as $\sigma(X)$. We require $X$ to be as follows.
It is not compact, but the noncompactness comes only from 
a cylindrical region of the form $\bR_{\geq} \times Y$ where $Y$ is compact. The metric in this region is assumed to be
a product of the metrics in $\bR_{\geq} $ and $ Y$. In the case of sigma models, 
the supercharge $Q$ is a kind of Dirac operator on $Z$.
The discussion of this section is really motivated by the analogy between Dirac operator and supercharge. 
For string-manifolds, mathematical invariants corresponding to the torsion index have been discussed in \cite{Bunke}. 

Another class of examples is as follows. Let $x$ be the quantum mechanical mode in the region $\bR_{\geq} =\{ x \in \bR~|~ x \geq 0\}$. 
Suppose that $x$ is actually defined as an elementary scalar field not just for $\bR_{\geq }$ but for the entire $\bR$,
but its potential $V(x)$ behaves as $V(x) \to x^2$ at $x \to -\infty$. 
If $V(x) = x^2$ in the entire region $x \in \bR$, then it would be a massive field and the energy spectrum satisfies
the condition of compactness. On the other hand, if $V(x) \to x^2$ at $x \to -\infty$ and $V(x) \to 0$ at $x \to \infty$,
then it is expected to be mildly noncompact.

In general, we expect to have state vectors $\ket{x,a}$, where $x \in \bR_{\geq}$ 
is the position of the quantum mechanical mode in the region $x>0$
(which is the zero mode of the boson of the sigma model $\sigma(\bR_{\geq}) $ in the space $S^1$),
and $a$ is any other discrete label specifying the quantum state.
It is normalized in the usual way, $\bra{x,a} x',b\rangle = \delta(x-x')\delta_{ab}$.
These states are expected to be well-defined for $x > 0$ because they represent states
localized at $x$ and hence they do not see the black (compact) region in Fig.~\ref{fig:nonC}. 
Then, for a given state vector $\ket{\Psi}$, 
we can consider its wave function $\Psi_a(x) = \bra{x,a}\Psi \rangle$ for $x>0$.
However, we remark that this wave function describes only partial information of $\ket{\Psi}$
since the wave function is defined only in the region $x>0$. 
Two different quantum states $\ket{\Psi} \neq \ket{\Psi'}$ can have the same wave function $\Psi_a(x) = \Psi'_a(x)$
in the region $x>0$ if their difference is only in the interior region.

Let $\cZ$ be a mildly noncompact SQFT with a boundary theory $\cY$. 
We are going to define an index $I_\cZ(q)$ of $\cZ$. The eigenvalues of the momentum operator $P$
take the form $n-\nu/24$, where $\nu \in \bZ \simeq (\IOspin)^4(\pt)$ is the pure gravitational anomaly of $\cZ$
with the sign convention that a sigma model $\sigma(Z)$ has $\nu =  \dim Z$ coming from right moving fermions $\psi^I~(I=1,\cdots,\dim Z)$.\footnote{
This shift of momentum may be argued as follows. If there were no anomaly, then $\exp(2\pi \i P)$ is a trivial translation on $S^1$
and hence should be the identity. Thus the nontrivial values of $\exp(2\pi \i P)$ should be due to gravitational anomalies.
Now we may add decoupled spectator free fermions so that the total anomaly is zero. After adding them, the total system
has $\exp(2\pi \i P)=1$. Thus the computation of $\exp(2\pi \i P)$ is reduced to that of free fermions. 
}
We decompose the Hilbert space $\cH$ of the theory $\cZ$ in terms of the eigenvalues of $P$ 
\beq
\cH = \bigoplus_{n \in \bZ} \cH_{n-\nu/24}, \qquad P \cH_{n-\nu/24} = (n-\nu/24) \cH_{n-\nu/24}.
\eeq
We consider the case that $(-1)^F$ is defined on $\cH$, which requires that $\nu$ is even.

Now we want to consider kernels of the supercharge $Q$ in the Hilbert space.
However, we can consider two types of kernels. One type, which we denote as $\Ker Q$,
consists of states $\ket{\Psi} \in \cH$ such that $Q\ket{\Psi}=0$ and $\ket{\Psi}$ is normalizable, $\bra{\Psi} \Psi \rangle < \infty$. 
In the noncompact region $x > 0$, we can consider wave functions $\Psi_a(x) =\bra{x,a} \Psi\rangle$.
In the definition of $\Ker Q$, the condition $\bra{\Psi} \Psi \rangle < \infty$ requires that 
$\Psi_a(x)$ decay first enough so that $\sum_a \int \d x   |\Psi_a(x)|^2 < \infty$.
Another type of kernel, which we denote $\Kerp Q$, is defined by requiring the wave functions
to be bounded as $\sum_a |\Psi_a(x)|^2 < \textrm{(const.)}$ for $\forall x >0$.
We do not require them to be square normalizable. 

Let us see more explicitly the behavior of the wave functions in the region $x \to \infty$.
The supercharge $Q$ in the region $x > 0$ is of the following form.
Let $x$ and $\gamma_x$ be the zero modes of the boson and the fermion of the sigma model $\sigma(\bR_{\geq})$.
In particular, we have $\gamma_x^2=1$ and $\gamma_x$ anticommutes with any other fermion modes.
Notice that $x$ and $\gamma_x$ are superpartners of each other. 
Let $p_x =- \i \frac{\partial}{\partial x}$ be the momentum operator corresponding to $x$.
Then the supercharge is given by
\beq \label{eq:Qcylinder}
Q =  \gamma_x p_x + Q',
\eeq
where $Q'$ is constructed from modes other than $x$ and $\gamma_x$, including nonzero modes of the sigma model $\sigma(\bR_{\geq})$.
More precise statement is that we can consider
matrix elements $\bra{x,a} Q \ket{\Psi}$ and these matrix elements are expressed as \eqref{eq:Qcylinder} for $x>0$.
Let $\widehat{Q}$ be the operator
\beq\label{eq:bdyQ}
\widehat{Q} = \i \gamma_x Q'.
\eeq
This operator is self-adjoint since $Q'$ and $\gamma_x$ are self-adjoint and anticommute with each other.
Also, $\widehat{Q}$ does not contain the noncompact mode $x$. 
Thus we expect $\widehat{Q}$ to have a real discrete spectrum. Suppose that its eigenvalue on $\ket{x,a}$
is $\lambda_a \in \bR$, 
\beq
\widehat{Q} \ket{x,a} = \lambda_a \ket{x,a}.
\eeq
If the state $\ket{\Psi}$ is in the kernel of $Q$, 
the wave function $\Psi_a(x)$ behaves as follows:
\beq\label{eq:wavebehavior}
&0 = Q\Psi_a(x) = -\i \gamma_x \left( \frac{\partial}{\partial x} + \lambda_a \right)\Psi_a(x) \nonumber \\ 
&\Longrightarrow \quad \Psi_a(x) \propto e^{-\lambda_a x}
\eeq
For the state $\ket{\Psi}$ to be in $\Ker Q$,
we must have $\Psi_a(x)=0$ for all $a$ such that $\lambda_a \leq 0$. 
On the other hand, for $\ket{\Psi}$ to be in $\Kerp Q$,
we must have $\Psi_a(x)=0$ for all $a$ such that $\lambda_a < 0$. 
This is the difference between $\Ker Q$ and $\Kerp Q$.
If all eigenvalues of $\widehat{Q}$ are nonzero, i.e. $\lambda_a \neq 0~(\forall a)$, then $\Ker Q=\Kerp Q$.

We denote by $(\Ker Q)^{\pm}_{n-\nu/24}$ and $(\Kerp Q)^{\pm}_{n-\nu/24}$
the subspaces of $\Ker Q$ and $\Kerp Q$ with the eigenvalues of $(-1)^F$ and $P$ given by
$\pm 1$ and $n-\nu/24$, respectively. These spaces are expected to be finite dimensional by the following reason.
In these spaces, we have $Q^2=\frac{1}{2} (H-P)=0$ and hence $H=P=n-\nu/24$, so the states have a fixed energy. 
In the situation as in Fig.~\ref{fig:nonC} in which the noncompact direction is one-dimensional, 
we expect to have only finite number of states at a given energy.

We remark that states in $(\Kerp Q)^{\pm}_{n-\nu/24}$ with $\lambda_a=0$ may be part of a continuous spectrum
of states whose wave functions are of the form $\Psi(x) \sim A e^{-\si k x} + Be^{\si k x}$ for $k \in \bR$.
These are states which are injected from the region $x = \infty$ with momentum $p_x = -k$,
reflected by the compact region, and then going out to the region $x = \infty$ with momentum $p_x = k$.
The eigenvalue of $Q^2$ is $k^2$, and $k=0$ gives a state in $\Kerp Q$.
But this does not change the conclusion about the finiteness of $(\Kerp Q)^{\pm}_{n-\nu/24}$
since the number of wavefunctions with $k=0$ is still expected to be finite.

We define
\beq\label{eq:ind1}
\index (Q_{n-\nu/24}) =&~ \dim (\Kerp Q)^{+}_{n-\nu/24} - \dim (\Ker Q)^{-}_{n-\nu/24} \nonumber \\
\indexp (Q_{n-\nu/24}) =&~ \dim (\Ker Q)^{+}_{n-\nu/24} - \dim (\Kerp Q)^{-}_{n-\nu/24} 
\eeq
and
\beq\label{eq:Iindex}
I_\cZ(q) = &~\sum_{n \in \bZ} \index (Q_{n-\nu/24} ) q^{n-\nu/24} \nonumber \\
\widetilde{I}_\cZ(q) = &~\sum_{n \in \bZ} \indexp (Q_{n-\nu/24} ) q^{n-\nu/24} 
\eeq
where $q = e^{2\pi \i \tau}$ is a variable. 
Notice that $ \index (Q_{n-\nu/24} )$ and $\indexp (Q_{n-\nu/24})$ are zero for sufficiently negative $n$.
The reason is that in the kernel of $Q$
we have $P = H$, but $H$ should be bounded from below.
Also notice that if $\cZ$ is a compact SQFT, then we have $I_\cZ(q) = \tr (-1)^F q^{H_L} \bar{q}^{H_R}$ and it is the
elliptic genus~\cite{Witten:1986bf}.

The index \eqref{eq:ind1} is defined by imitating the definition of the 
APS index of Dirac operators~\cite{Atiyah:1975jf}.
In fact, for sigma models $\cZ=\sigma(Z)$, the relation between $I_\cZ(q)$ and the APS index in mildly noncompact theories
is the same as the relation between the elliptic genus and the Atiyah-Singer index in compact theories~\cite{Witten:1986bf}.
The motivation for using both $\Ker$ and $\Kerp$ in the definition \eqref{eq:ind1}
is the same as the case of the APS index, and we will explain it below.

The index $I_\cZ(q)$ has an important gluing law. To explain it, let us first
introduce the notion of orientation reversal $\overline{\cZ}$ of an SQFT $\cZ$. 
It is just defined by adding a topological term corresponding to the nontrivial element of
$\Hom(\Omega^{\spin}_2(\pt), \bR/\bZ) \simeq \bZ_2$. (See also Sec.~\ref{sec:Bfield} for related discussions.)
This is a generalized theta term, and its effect is that if the worldsheet has the odd spin structure (meaning that the mod 2 index is nontrivial),
then this theta angle assigns an additional phase $(-1)$ to the partition function. 
In particular, the action of $(-1)^F$ on Ramond sector states gets an additional factor $(-1)$.
Thus the index is changed as
\beq
I_{\overline{\cZ}}(q) = - \widetilde I_{\cZ}(q), \qquad \widetilde I_{\overline{\cZ}}(q) = - I_{\cZ}(q),
\eeq

In the noncompact region, the theory is of the form $\sigma(\bR_{\geq}) \otimes \cY$.
If we flip the $\bR_{\geq}$ direction, then the fermion $\psi$ in the chiral multiplet is transformed as $\psi \to -\psi$.
This transformation is anomalous and the theta angle mentioned above is produced.
Thus, in the noncompact region, the effect of the orientation reversal can be compensated by the 
change of the direction of $\bR_{\geq}$.

\begin{figure}
\centering
\includegraphics[width=0.7\textwidth]{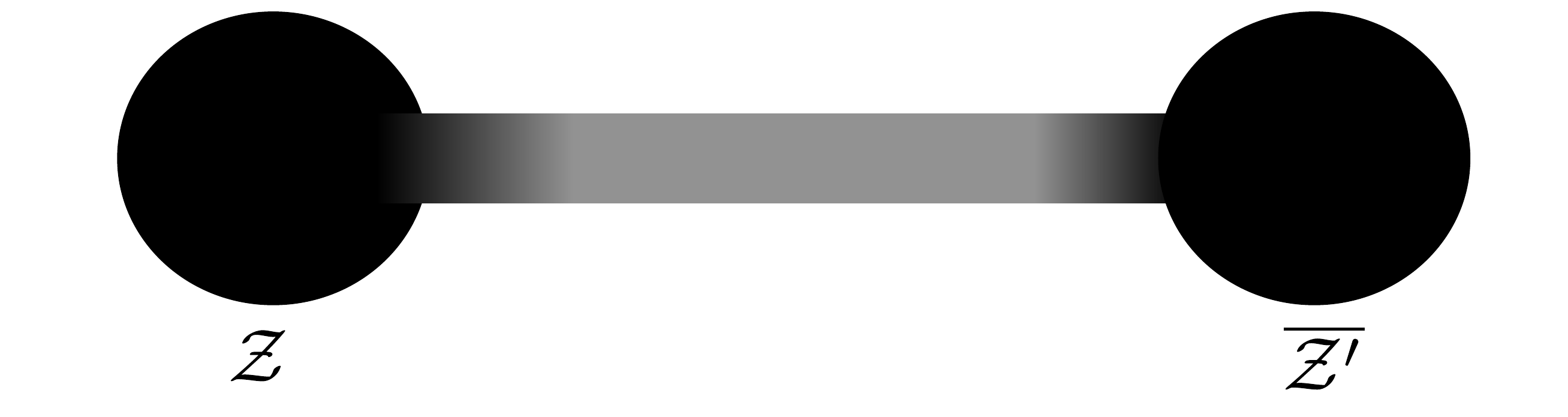}
\caption{Gluing of two mildly noncompact SQFTs $\cZ$ and $\cZ'$. The orientation of $\cZ'$ is reversed to $\overline{\cZ'}$. 
 \label{fig:glue}}
\end{figure}

If we have two mildly noncompact SQFTs $\cZ$ and $\cZ'$ with the same boundary theory $\cY$,
we can glue $\cZ$ and $\overline{\cZ'}$ as in Fig.~\ref{fig:glue}. 
This gluing is possible because the orientation reversal
on $\cZ'$ changes the direction of $\bR_{\geq}$ in the noncompact region.
For example, in sigma models this gluing is just the gluing of two manifolds along their common boundaries.
However, our construction applies to any mildly noncompact SQFTs. 

By gluing  $\cZ$ and $\cZ'$, we obtain a compact SQFT $\cZ''$.
Now we claim that the index $I_{\cZ}(q)$ satisfies the gluing law
\beq\label{eq:gluinglaw}
I_{\cZ''}(q) = I_{\cZ}(q) + \widetilde I_{\overline{\cZ'}}(q) =I_{\cZ}(q) - I_{\cZ'}(q).
\eeq
This is argued as follows. (The following arguments are borrowed from the corresponding arguments for
the APS index of Dirac operators.)

First let us consider the case that all eigenvalues of $\widehat{Q}$ defined in \eqref{eq:bdyQ} are nonzero. 
Then, $Q^2 = p_x^2 + \widehat{Q}^2$ and $\widehat{Q}^2>0$. In the theory $\cZ$,
the positive definiteness $\widehat{Q}^2>0$ means that we need a positive ``energy'' to go to the region $x \to \infty$,
where the ``energy'' is in the sense of $H_R=Q^2$.
Thus, all the almost zero energy states of $H_R=Q^2$ are localized in the black compact region of Fig.~\ref{fig:nonC}.
When we glue two theories $\cZ$ and $\cZ'$, we can take the cylindrical region $\sigma(\bR) \otimes \cY$
to be very long. Then, almost zero energy states must be localized in the black compact region
of either $\cZ$ or $\overline{\cZ'}$ in Fig.~\ref{fig:glue}. 
Because of the deformation invariance of the index~\cite{Witten:1982df}, 
the index can be computed by counting almost zero energy states without caring whether the energy is exactly zero or not.
Thus the index is unchanged before and after the gluing, and we get the desired gluing law.
Notice that if $\widehat{Q}$ does not have zero eigenvalues, then $\Ker Q = \Kerp Q$ and hence $I_{\cZ}(q) = \widetilde I_{\cZ}(q)$.

Next we want to consider the case that $\widehat{Q}$ may have zero eigenvalues. 
In this case, we can argue as follows.
Let $\epsilon>0$ be a sufficiently small constant such that for all nonzero eigenvalues $\lambda_a \neq 0 $ of $\widehat Q$,
we have $\epsilon < |\lambda_a|$. 
In the theory $\cZ$, we consider the following deformed operator $ Q_\epsilon$.
Let $\rho(x)$ be a function such that
\beq
\rho(x) = \left\{ \begin{array}{ll}
\epsilon x & (x \to \infty) \\
\textrm{const.} & (x <0)
\end{array}\right.
\eeq
By using it, we define a self-adjoint operator
\beq  \label{eq:deformQ}
 Q_\epsilon := e^{(-1)^F \rho(x)} Q e^{(-1)^F \rho(x)}.
\eeq
This is defined in the entire region because $\rho(x)$ is taken to be just constant in the region $x<0$.
The kernels of $Q$ and $  Q_\epsilon$
are related by simply acting $e^{-(-1)^F \rho(x)}$ to states. More precisely, we have 
\beq
&(\Kerp Q)^+ \ni \ket{\Psi_+} \mapsto e^{- \rho(x) } \ket{\Psi_+} \in (\Ker  Q_\epsilon )^+, \nonumber \\
&(\Ker Q)^- \ni \ket{\Psi_-} \mapsto e^{+ \rho(x) } \ket{\Psi_-} \in (\Ker  Q_\epsilon )^-,
\eeq
where $(-1)^F \ket{\Psi_\pm } = \pm  \ket{\Psi_\pm } $.
We can see that both $e^{- \rho(x) } \ket{\Psi_+}$ and $e^{ \rho(x) } \ket{\Psi_-}$ 
for $ \ket{\Psi_+} \in (\Kerp Q)^+$ and $\ket{\Psi_-} \in (\Ker Q)^- $
decay exponentially at $x \to \infty$ because of our choice of $\epsilon$. 
These maps give isomorphisms
\beq\label{eq:keriso}
(\Kerp Q)^+ \simeq (\Ker  Q_\epsilon )^+,  \qquad   (\Ker Q)^- \simeq    (\Ker Q_\epsilon )^-.
\eeq
In fact, in the region $x \to \infty$, we have
\beq
\widetilde Q = -\i \gamma_x \left(  \frac{\partial}{\partial x} + \widehat{Q} + \epsilon (-1)^F  \right).
\eeq
To obtain this formula, we have used the fact that $(-1)^F$ anticommutes with $\gamma_x$.
The only difference between $Q$ and $Q_\epsilon $ is that $\widehat{Q}$ is replaced by $\widehat{Q} + \epsilon (-1)^F $.
The operator $\widehat{Q} + \epsilon (-1)^F$ does not have zero eigenvalue because of our choice of $\epsilon$,
and
\beq
\lambda_a \geq 0 \Longleftrightarrow \lambda_a + \epsilon >0 , \qquad \lambda_a >0 \Longleftrightarrow \lambda_a - \epsilon >0.
\eeq
Therefore, by the behavior of wavefunctions \eqref{eq:wavebehavior}, we get the desired isomorphisms.

We can use the operator $ Q_\epsilon $ 
for the definition of the index because of the isomorphisms \eqref{eq:keriso}.
Since $\widehat{Q} + \epsilon (-1)^F$ does not have zero eigenvalue, the previous result applies to $Q_\epsilon$
and we get the desired formula \eqref{eq:gluinglaw}. 

Now we can understand why two types of kernels $\Ker$ and $\Kerp$ are used in the definition \eqref{eq:ind1}.
We have the isomprhisms \eqref{eq:keriso} after the deformation \eqref{eq:deformQ},
and the definition of the index in terms of $Q_\epsilon$ for $\epsilon>0$ gives the index as $\dim\Kerp Q - \dim\Ker Q$.
We could also consider another deformation by replacing $\epsilon \to -\epsilon$.
In that case, we get isomorphisms $(\Ker Q)^+ \simeq (\Ker  Q_{-\epsilon} )^+$ and $ (\Kerp Q)^- \simeq    (\Ker Q_{-\epsilon} )^-$
and hence the corresponding index is given by $\dim\Ker Q - \dim\Kerp Q$. Thus we also have another gluing law
$I_{\cZ''}(q)  =\widetilde I_{\cZ}(q) - \widetilde I_{\cZ'}(q)$.

Incidentally, the relation between $ I_{\cZ}(q)$ and $\widetilde I_{\cZ}(q)$ can be derived
by using the gluing law. Let $\cC = \sigma(\bR) \otimes \cY$.
We decompose the wave function as $\Psi_a(x) = \Psi^+_a(x) + \Psi^-_a(x)$ where $ \Psi^+_a(x) $ is the part with $(-1)^F=+1$
and $\Psi^-_a(x)$ is the part with $(-1)^F=-1$.
We impose the boundary condition that $ \Psi^+_a(x) $ is bounded for $\forall x \in \bR$, and $ \Psi^-_a(x) $ is square-normalizable.
Let $I_\cC(q)$ be the index with this boundary condition for $\cC$.
If we glue the region $x\to \infty$ of $\cZ$ and the region $x \to -\infty$ of $\cC$, 
we again get $\cZ$. However, the boundary condition is different before and after the gluing.
Before the gluing, we need to impose the boundary condition for $\cZ$ such that the index is $\widetilde I_{\cZ}(q)$.
This is necessary so that the gluing of $\cZ$ and $\cC$ is possible. After the gluing, the boundary condition
is such that the index is $ I_{\cZ}(q)$.
By the gluing law, we get
\beq
I_{\cZ}(q) = \widetilde I_{\cZ}(q) + I_{\cC}(q).
\eeq
One can see that $I_{\cC}(q)$ is given in terms of $\cY$ as follows.
The Hilbert space of $\cY$ is decomposed as 
\beq
\cH^\cY= \bigoplus_{n \in \bZ} \cH^\cY_{n-(\nu-1)/24},
\eeq
where $\cH^\cY_{n-(\nu-1)/24}$ is the subspace with $P =  n-(\nu-1)/24$.
Let $\Ker Q^\cY_{n-(\nu-1)/24}$ be the kernel of the supercharge $Q^\cY$ in the subspace $\cH^\cY_{n-(\nu-1)/24}$.
Then we have
\beq
I_{\cC}(q) = \eta(\tau)^{-1} \sum_{n \in \bZ} q^{n-(\nu-1)/24} \dim \Ker Q^\cY_{n-(\nu-1)/24}
\eeq
where $\eta(\tau)=q^{1/24}\prod_{n\geq 1}(1-q^n)$ is the Dedekind $\eta$-function. 
The contribution of $\eta(\tau)$ comes from the left-moving excited modes of $\sigma(\bR)$ in $\cC = \sigma(\bR) \otimes \cY$.

\subsection{The torsion index}\label{sec:TI}
We have discussed some properties of mildly noncompact SQFTs.
Now we can give a definition of the torsion index. 

To define the torsion index of an SQFT $\cY$, we assume the following.
\begin{enumerate}
\item $\cY$ is a compact SQFT with the pure gravitational anomaly $\nu(\cY)=\nu-1$.
We assume $\nu$ is even, so that the theory $\sigma(\bR) \otimes \cY$ 
has the even pure gravitational anomaly $\nu$ 
and hence $(-1)^F$ is defined.
\item There exists an integer $N$ and an SQFT $\cZ$ such that $N$ copies $\cY^{\oplus N} = \cY \oplus \cdots \oplus \cY$ 
of the theory $\cY$ is the boundary theory of $\cZ$. (Only the existence of $N$ and $\cZ$ is required, and we will argue
that the torsion index does not depend on their choice.) 
\end{enumerate}
Then the torsion index, which we denote as $J_\cY(q)$, will be defined by
\beq
J_\cY(q) = \frac{1}{N} I_{\cZ}(q) \quad \textrm{modulo some quantities}
\eeq
where $ I_{\cZ}(q)$ is the index of $\cZ$ defined in Sec.~\ref{sec:noncompact}. 
We want $J_\cY(q)$ to be independent of the choice of $N$ and $\cZ$,
and we also want $J_\cY(q)$ to be invariant under deformation of $\cY$ 
(in the sense of ``flowing up and down the renormalization group trajectories'' as discussed in \cite{Gaiotto:2019asa}).
Therefore we will need to divide by some quantities.

First let us discuss the dependence on $(N, \cZ)$. 
Suppose that $N'$ copies of $\cY$ can be realized as the boundary theory of $\cZ'$.
Then, both $(\cZ)^{\oplus N'}$ and $(\cZ')^{\oplus N}$ has $\cY^{\oplus NN'}$ as the boundary theory,
so we can glue them to get a compact theory $\cZ''$. By the gluing law \eqref{eq:gluinglaw}, we get
\beq
 \frac{1}{N} I_{\cZ}(q) -  \frac{1}{N'} I_{\cZ'}(q) = \frac{1}{NN'} I_{\cZ''}(q).
\eeq
Let $\eta(\tau)=q^{1/24}\prod_{n\geq 1}(1-q^n)$ be the Dedekind $\eta$-function. 
For a compact theory $\cZ''$ with the gravitational anomaly $\nu$, 
the Witten genus $\eta(\tau)^{\nu} I_{\cZ''}(q)$ is a weakly holomorphic modular form of weight $\nu/2$.
Here, ``weakly holomorphic'' means that we can have poles at $q=0$.
We denote by $\mathrm{MF}[\Delta^{-1}]_{n}$ the set of weakly holomorphic modular forms of weight $n$
(where $\Delta=\eta(\tau)^{24}$ and this notation indicates that we can have powers of $\Delta^{-1}=q^{-1} +\cdots$).
More explicitly, if $f(\tau) \in \mathrm{MF}[\Delta^{-1}]_{n}$, then $f(\tau) $ is a Laurent series of $q=e^{2\pi \i \tau}$,
$f(\tau+1)=f(\tau)$ and $f( - 1/\tau)= \tau^n f(\tau)$.

We conclude
\beq
 \frac{1}{N} I_{\cZ}(q) -  \frac{1}{N'} I_{\cZ'}(q) \in \eta(\tau)^{-\nu} \mathrm{MF}[\Delta^{-1}]_{\nu/2} \otimes \bQ,
\eeq
where $\bQ$ is the set of rational numbers. 
Thus, $ \frac{1}{N} I_{\cZ}(q) $ is independent of the choice of $(N, \cZ)$ modulo $\eta(\tau)^{-\nu} \mathrm{MF}_{\nu/2} \otimes \bQ$.

Next let us discuss deformation of $\cY$. If necessary, we embed $\cY$ to a larger UV theory as discussed in \cite{Gaiotto:2019asa}
such that the low energy theory is unchanged up to irrelevant operators. 
Suppose that there is a one parameter family $\{\cY_s\}$
of SQFTs parametrized by $s \in [0,1]$. We want to make sure that $\cY_0$ and $\cY_1$ have the same torsion index.

Let $\rho(x)$ be a function of $x \in \bR$ such that 
\beq
\rho(x) \to \left\{ \begin{array}{ll}  
0 &  (x \to -\infty) \\
1 & (x \to +\infty)
\end{array}\right. .
\eeq
Then, we introduce a dynamical chiral multiplet $\Phi$ in such a way that its scalar component $\phi$
gives the deformation of $\cY_s$ as $s = \rho(\phi)$. Then we obtain a new mildly noncompact SQFT $\cW$ 
whose gravitational anomaly is $\nu$. Let $I_{\cW}(q)$ be the index of $\cW$ with the boundary condition that
positive chirality wave functions $\Psi_a(x),~(-1)^F\ket{x,a}=\ket{x,a}$ are bounded at $x \to +\infty$ and decay exponentially at $x \to -\infty$,
while negative chirality wave functions $\Psi_a(x),~(-1)^F\ket{x,a}=-\ket{x,a}$ 
decay exponentially at $x \to +\infty$ and are bounded at $x \to -\infty$. 

If we are given a theory $\cZ_0$ whose boundary theory is $\cY_{0}^{\oplus N}$,
we can construct a theory $\cZ_1$ whose boundary theory is $\cY_{1}^{\oplus N}$ by
gluing $\cW^{\oplus N}$ to $\cZ_0$.
Then, by a similar gluing argument to that in Sec.~\ref{sec:noncompact}, we get
\beq
\frac{1}{N}I_{\cZ_1}(q) - \frac{1}{N}I_{\cZ_0}(q) = I_{\cW}(q).
\eeq
By definition, the index $ I_{\cW}(q)$ is an element of $q^{-\nu/24} \bZ((q))$, 
where $\bZ((q))$ is the set of Laurent series of $q$ with integer coefficients. 
Thus the torsion index $J_{\cY}(q)$ is invariant under deformation modulo $q^{-\nu/24} \bZ((q))$. 

We can obtain a stronger result when $\nu \equiv 4 \mod 8$.
The CPT symmetry $\mathsf{CPT}$ of a theory with a pure gravitational anomaly $\nu$ has the following properties:
\beq
\begin{array}{ll}
\mathsf{CPT}^2= -1, \quad \mathsf{CPT} (-1)^F =(-1)^F \mathsf{CPT}   &\qquad (\nu \equiv 4 \mod 8) \\
 \mathsf{CPT} (-1)^F + (-1)^F \mathsf{CPT} =0 & \qquad (\nu \equiv 2 \mod 4) 
\end{array}
\eeq
One can check these statements explicitly for free Majorana-Weyl fermions in which case they follow from the properties of the Clifford algebra
of zero modes. Then they are true for any theory because anomalies are universal.\footnote{
More concretely, one can add decoupled Majorana-Weyl fermions to a theory to cancel the gravitational anomaly.
Then $\mathsf{CPT}$ and $(-1)^F$ satisfy the non-anomalous relation $\mathsf{CPT}^2= 1, \quad \mathsf{CPT} (-1)^F =(-1)^F \mathsf{CPT}  $, and we can consider the effect of $\mathsf{CPT}$ and $(-1)^F$
to each decoupled theories. See also \cite{Delmastro:2021xox} for related discussions.}
When $\nu =4 \mod 8$, the existence of the antiunitary operator $\mathsf{CPT}$ with $\mathsf{CPT}^2=-1$
implies the 2-fold degeneracy of states (i.e. Kramers doubling).
Thus, we have $I_{\cW}(q) \in 2q^{-\nu/24} \bZ((q))$ in this case.
For $\nu=2 \mod 4$, the operator $\mathsf{CPT}$ relates positive and negative chirality states $(-1)^F=\pm 1$
because of the anticommutation relation between $\mathsf{CPT}$ and $(-1)^F$.
For compact SQFTs, this implies that the index is zero. However, for noncompact SQFTs,
the definition of the index is asymmetric between positive and negative parts,
so the index $I_{\cW}(q)$ may not necessarily vanish.\footnote{
This non-vanishing of $\cI_{\cW}(q)$ for $\nu \equiv 2 \mod 4$ is of the following nature. 
If the boundary operator $\widehat{Q}$ defined in \eqref{eq:bdyQ} does not have zero eigenvalue,
then the difference of the boundary conditions does not matter and $\cI_{\cW}(q)=0$. Thus,
$\cI_{\cW}(q)$ can be nonzero only if $\widehat{Q}$ has zero modes. For example, suppose
that the family of theories $\cY_s$ does not generically have zero eigenvalues for $Q$.
However, at some points in the parameter space $s \in [0,1]$, we may get an accidental zero of $Q$.
In such a case, a spectral flow consideration as in the case of a family of Dirac operators as well as the boundary conditions 
may show that $\cI_{\cW}(q)$ can be nonzero if $\cY_0$ or $\cY_1$
is such an accidental point. 
}

In conclusion, we define the torsion index $J_{\cY}(q)$ of a theory $\cY$ with an odd gravitational anomaly $\nu(\cY)=\nu-1$ as
\beq\label{eq:deftorsion}
J_{\cY}(q) =  \frac{1}{N} I_{\cZ}(q)  \mod ~   \eta(\tau)^{-\nu} \mathrm{MF}[\Delta^{-1}]_{\nu/2} \otimes \bQ  + \mathsf{m} q^{-\nu/24} \bZ((q)),
\eeq
where $\cZ$ is a mildly noncompact SQFT whose boundary theory is $\cY^{\oplus N}$, 
$I_{\cZ}(q)$ is the index of $\cZ$ defined in Sec.~\ref{sec:noncompact}, and
\beq
\mathsf{m} = \left\{ \begin{array}{ll} 
2 \quad & (\nu \equiv 4 \mod 8) \\   
1 \quad & (\nu \equiv 0,2,6 \mod 8)
\end{array}\right. .
\eeq
We will give simple examples later in Sec.~\ref{sec:anomaly}.

Let us discuss the implication of the torsion index for spontaneous supersymmetry breaking in the infinite volume.
Suppose that $N$ copies of $\cY$ can be deformed to break supersymmetry 
(by embedding it into a larger UV theory if necessary). 
More specifically, we assume that there is a one parameter family of theories $ \cY_{s}$ parametrized by $s \in [0,\infty)$
such that the vacuum energy density (i.e. cosmological constant) in the infinite volume is given by $s $ (at least for large $s$),
and $ \cY_{0} = \cY^{\oplus N}$. 
In this situation, we can construct an SQFT $\cZ$ by introducing a chiral multiplet $\Phi$
and taking its scalar component $\phi$ to be $s = \rho(\phi)$, where 
\beq
\rho(\phi) \to \left\{
\begin{array}{cc}
\phi^2  & (\phi \to -\infty) \\
 0 & ( \phi \to +\infty).
\end{array}\right.
\eeq
The region $\phi \to -\infty$ has the energy density $\phi^2$ that grows like the potential energy of a massive scalar. 
Thus, this region is expected to be compact in the sense of the energy spectrum. Therefore,
we get a theory $\cZ$ with the boundary $\cY^{\oplus N}$ at $\phi \to + \infty$. 
In particular, if $N=1$, we get $J_{\cY}(q) = I_{\cZ}(q) $ which is $0 $ modulo $ \mathsf{m} q^{-\nu/24} \bZ((q))$.
Therefore, theories which can be deformed to break supersymmetry has the trivial torsion index. 
In other words, if an SQFT has a nontrivial value of the torsion index, 
supersymmetry cannot be spontaneously broken by a free parameter $s$ of the theory.
We remark that it is possible that $S^1$ compactification breaks supersymmetry 
by a vacuum energy of order the inverse radius of $S^1$~\cite{Gaiotto:2019asa}.

It was conjectured \cite{Stolz:2004,Stolz:2011zj} that the space of $\cN=(0,1)$ SQFTs with a worldsheet pure gravitational anomaly 
$\nu \in \bZ  $ is homotopy equivalent
to the $(-\nu)$-th space $\mathrm{TMF}_{-\nu}$ of a generalized cohomology theory known as topological modular forms $\mathrm{TMF}$.
This conjecture implies, 
among other things, that the cohomology group of a point $\mathrm{TMF}^{-\nu}(\pt) = \pi_0( \mathrm{TMF}_{-\nu})$ 
is the group of deformation classes of SQFTs with the gravitational anomaly $\nu$. Here, two SQFTs $\cY_0$ and $\cY_1$ 
are deformation-equivalent if they can be connected by a one-parameter family of theories $\{\cY_s\}~(s \in [0,1])$.
The zero element $0 \in \mathrm{TMF}^{-\nu}(\pt)$ may be represented by an SQFT with spontaneous supersymmetry breaking. 
(See \cite{Johnson-Freyd:2020itv} for more discussions.) 
The torsion index gives some of the invariants suggested by $\mathrm{TMF}^{-\nu}(\pt)$. 
If the torsion index of a theory $\cY$ with the gravitational anomaly $\nu(\cY)$ is nonzero, 
its deformation class $[\cY] \in \mathrm{TMF}^{-\nu(\cY)}(\pt)$ should represent a nontrivial element of $\mathrm{TMF}^{-\nu(\cY) }(\pt)$,
assuming the conjecture is true.

\section{Global anomalies and the torsion index}\label{sec:anomaly}
In this section, we study global anomalies of heterotic string theories.
However, we do not assume that the worldsheet SQFT is conformal nor has the correct gravitational anomaly.
We just study SQFTs without integrating over the worldsheet supergravity. We will see that target space anomalies
are rephrased as the index of a class of SQFTs. 

Because we consider general gravitational anomaly of the worldsheet, it is possible that the target space has some anomalies.
As we discuss, perturbative anomalies of the target space theory are represented by the usual Witten index,
and global anomalies are represented by the torsion index.

\subsection{Target space anomalies and index of SQFTs}

The basic setup we consider is the one described in Sec.~\ref{sec:BW}.
Let us briefly recall it. We consider an internal SQFT $\cT$ with a global symmetry $G$ and a parameter space 
$\cP$ which is regarded as a background sigma model for $\cT$.
We also consider a sigma model with a target space $X$. Here $X$ is assumed to be equipped with 
a Riemannian metric, a $G$-bundle with a connection, and a map $X \to \cP$.
Then $\cT$ is coupled to the sigma model $X$ via the connection of the $G$-bundle and the map $X \to \cP$.
For this coupling between $\cT$ and $X$ to be well-defined, 
the anomaly cancellation condition discussed in Sec.~\ref{sec:BW} must be satisfied.
We denote the combined theory as $\sigma(X,\cT)$.
We also denote the pure gravitational anomaly of a theory $\cX$ as $\nu(\cX)$, and in particular
we have $\nu(\sigma(X,\cT)) = \dim X + \nu(\cT)$.

In the following discussion, we always assume that a target space $X$ is large enough so that weakly coupled description
for the sigma model is possible. ($\cT$ is general and may be strongly coupled.) We also assume that
$\nu(\sigma(X,\cT)) $ is even so that the GSO projection to states with $(-1)^F=+1$ makes sense. 
However, the GSO projection may not be compatible with the Majorana (i.e. real) structure imposed by $\mathsf{CPT}$.
We discuss the real structure a little more later. 

Let us recall how the target space theory is described.
It contains fermions and we are interested in their anomalies.\,\footnote{
We do not necessarily restrict our attention to massless fermions
to take into account the possibility of coupling space anomalies of the type discussed in \cite{Cordova:2019uob,Kanno:2021bze,Gomi:2021bhy,Choi:2022odr}.
When we vary parameters in $\cP$ of the theory $\cT$, some massive modes may become massless at some points of $\cP$
and they can produce anomalies. \label{foot:coupling}}
In actual heterotic string theories, we are only interested in the states with the worldsheet momentum $P=0$ (after including ghost fields and taking BRST cohomology). 
However, we consider all states with any value of $P$ unless otherwise stated.

By quantizing the theory except for the bosonic zero modes of the sigma model $X$, we get a space
\beq
\cS^{\sigma(X,\cT)}= \bigoplus_{n \in \bZ} \cS^{\sigma(X,\cT)} _{n-\nu(\sigma(X,\cT))/24}.
\eeq
where $ \cS^{\sigma(X,\cT)} _{n-\nu(\sigma(X,\cT))/24}$ is the eigenspace of $P$ with eigenvalue $P=n-\nu(\sigma(X,\cT))/24$,
and $ \cS^{\sigma(X,\cT)} _{n-\nu(\sigma(X,\cT))/24} =0$ for sufficiently negative $n$.
The symmetry $G$ acts on $\cS^{\sigma(X,\cT)}$.
This space is also a Clifford module because quantization of fermionic zero modes of the sigma model $X$ gives 
gamma matrices associated to the tangent bundle $TX$. 
Quantization of the bosonic zero modes of the sigma model $X$ gives quantum mechanical coordinates of $X$.
Then the space $\cS^{\sigma(X,\cT)}$ becomes a bundle (or more precisely a Clifford module of $TX$) over $X$, 
and the total Hilbert space $ \cH^{\sigma(X,\cT)}$ is the space of sections of the bundle $\cS^{\sigma(X,\cT)} $,
\beq
 \cH^{\sigma(X,\cT)} = \Gamma(\cS^{\sigma(X,\cT)} ).
\eeq
Here we have neglected winding modes in $X$ since we consider the large volume limit of $X$.
(It is possible that the internal theory $\cT$ is a small volume sigma model.)
The supercharge is of the form 
\beq\label{eq:QDM}
Q=\i\slashed{D} + \mathsf{M}, 
\eeq
where $\i \slashed{D}$ is the Dirac operator constructed from the bosonic and fermionic zero modes of the sigma model $X$,
and $ \mathsf{M}$ is constructed from other modes and can be interpreted as a mass term.
This $\mathsf{M}$ may depend on the parameters in $\cP$, and hence depend on the position in $X$ via the map $X \to \cP$.
Notice that $ \mathsf{M}$ anticommutes with $(-1)^F$, so it is odd under the $\bZ_2$-grading of $(-1)^F$.
We consider the target space fermions (including states with $P \neq 0$) described by this Dirac operator.
The operator $(-1)^F$ is interpreted as the target space chirality operator, and hence
the GSO projection $(-1)^F=+1$  corresponds to taking chiral fermions. 

Target space anomalies are classified by the Anderson dual of some bordism theory as discussed in Sec.~\ref{sec:review}.
For the current case, we are interested in target space manifolds $X$ equipped with a metric, a $G$-bundle with a connection, 
a map $X \to \cP$, and a $B$-field as discussed in Sec.~\ref{sec:BW}. 
At the topological level, the condition is given by \eqref{eq:topocondition}. 
We denote this type of structure (including all data mentioned above) on manifolds as $\mathcal{B}$, and the corresponding bordism groups as $\Omega^{\mathcal{B}}_\bullet$.
Thus the target space anomalies are classified by $(\IO^{\mathcal{B}})^{D+2}(\pt)$,
where $D = \dim X$ is the dimension of the target space.\footnote{Strictly speaking, 
the classification of anomalies in terms of $(\IO^{\mathcal{B}})^{\bullet}$ is not yet established in the literature for 
structure types like $\mathcal{B}$ which are sometimes called 2-groups.
We believe that this is not an essential problem and just assume that $(\IO^{\mathcal{B}})^{\bullet}$ is the correct cohomology theory
for the classification of anomalies.}
It satisfies the short exact sequence \eqref{eq:short} discussed in Sec.~\ref{sec:review},
\beq\label{eq:shortB}
0 \to \Ext(\Omega^{\mathcal{B}}_{D+1}(\pt), \bZ) \to (\IO^{\mathcal{B}})^{D+2}(\pt) \to \Hom(\Omega^{\mathcal{B}}_{D+2}(\pt), \bZ) \to 0
\eeq

Let us briefly recall how to describe anomalies of chiral fermions.
See \cite{Witten:2019bou} for detailed explanations of the following procedure.

Suppose that we have a chiral fermion $\chi_+$ in $D$-dimensions with positive chirality.
Let $\chi_-$ be the corresponding fermion with negative chirality.
Then we consider a $(D+1)$-dimensional massive fermion $\Psi$
such that if it is restricted to a manifold of the form $\bR \times X$,
then it has components of the form $(\chi_+, \chi_-)$. Putting this fermion on a manifold $Y$
with boundary $\partial Y = X$ with an appropriate boundary condition,
we get a localized chiral fermion $\chi_+$ on the boundary $X$. 
The APS $\eta$-invariant of a Dirac-like operator $\cD_{D+1}$ acting on this fermion $\Psi$ gives 
the $(D+1)$-dimensional bulk theory.\footnote{This Dirac-like operator $\cD_{D+1}$ may contain a mass term $ \mathsf{M}$
of the type discussed in \eqref{eq:QDM}. This mass is different from the mass that is used
to make the bulk $(D+1)$-dimensional theory completely gapped. We assume that the APS $\eta$-invariant
and the APS index theorem are still valid in the presence of such $\sM$ which is odd under the $\bZ_2$-grading. 
In our situation, $\sM$ may be an infinite dimensional matrix whose eigenvalues have spectral flows.
For instance, $\cT$ may be a sigma model $\sigma(S^1) \otimes \textrm{WZW}_k (\U(1))$ where $\textrm{WZW}_k (\U(1))$ is a level $k$
Wess-Zumino-Witten, and we may introduce a holonomy $\int_{S^1} a \in \bR/\bZ$ of a $\U(1)$ field $a$.
The holonomy is one of the parameters in $\cP$. The mass $\sM$ may be of the form $\sM \sim \diag (\cdots, n+\int_{S^1} a, \cdots)$
where $n \in \bZ$.
By changing the holonomy as $\int_{S^1} a \to \int_{S^1} a +1$, we get a situation
in which $\sM$ has a spectral flow. Presumably, the definition of the APS $\eta$-invariant and the APS index theorem work
even in such a situation by considering the $\eta$-invariant of the total self-adjoint operator $\cD_{D+1} = \i \slashed{D}+\sM$.
}
In the notation of Sec.~\ref{sec:review}, the bulk theory is given by 
\beq
(h,\omega) = (-\eta, \cI) \in \widehat {(\IO^{\mathcal{B}})}{}^{D+2}(\pt) 
\eeq
where $\eta$ is the APS $\eta$-invariant of $\cD_{D+1}$, 
and $\cI$ is the associated $(D+2)$-form that appears 
in the APS index theorem in $(D+2)$-dimensions. In more detail, the APS index of the corresponding Dirac operator $\cD_{D+2}$
on a $(D+2)$-manifold $Z$ with boundary $\partial Z$ is 
given by
\beq
\index \cD_{D+2}(Z)  = \int_Z \cI + \eta(\partial Z),
\eeq
where $\index \cD_{D+2}(Z)$ means the index of the Dirac operator $\cD_{D+2}$ on the manifold $Z$, and 
$ \eta(\partial Z)$ is the APS $\eta$-invariant of the Dirac operator $\cD_{D+1}$ on the boundary $\partial Z$.
This $(D+2)$-form $\cI$  is the anomaly polynomial of the chiral fermion $\chi_+$.
(We have not yet incorporated the Majorana condition. We will take it into account later.) 

Perturbative anomalies are described as follows. We consider the Atiyah-Singer index
on all possible closed $Z$ (i.e. $\partial Z = \varnothing$),
\beq\label{eq:pA}
\index \cD_{D+2}(Z) = \int_Z \cI.
\eeq
This gives a map
\beq
\Omega^{\mathcal{B} }_{D+2}(\pt) \ni [Z] \mapsto \index \cD_{D+2}(Z) \in \bZ.
\eeq
This is the part $\Hom(\Omega^\mathcal{B}_{D+2}(\pt), \bZ)$ in \eqref{eq:shortB}.
If this is zero for all elements of $\Omega^{\mathcal{B} }_{D+2}(\pt)$, 
the short exact sequence \eqref{eq:shortB} implies that we have $\cI= \d \cJ$
for some gauge invariant $(D+1)$-form $\cJ$. 
Then we can modify the bulk $(D+1)$-dimensional theory by adding a counterterm
$-(h_{\cJ}, \d \cJ)$, where as in Sec.~\ref{sec:review} we have defined
\beq
h_{\cJ}(Y) = \int_Y \cJ. 
\eeq
After adding the counterterm, we get
\beq\label{eq:Jcounter}
(\tilde h, 0): = (-\eta, \cI) -(h_{\cJ}, \d \cJ) .
\eeq
This is a new bulk theory which is equivalent to the original one in the sense discussed in Sec.~\ref{sec:review}. 
Now its anomaly polynomial is zero and $\tilde h$ is an element of $\Hom(\Omega^\mathcal{B}_{D+1}(\pt), \bR/\bZ)$.
The equivalence class $[(\tilde h,0)]$ is an element of 
\beq
\Ext(\Omega^\mathcal{B}_{D+1}(\pt), \bZ) = \Hom(\Omega^\mathcal{B}_{D+1}(\pt), \bR/\bZ)/  \Hom(\Omega^\mathcal{B}_{D+1}(\pt), \bR).
\eeq

Computation of the global anomaly $\tilde h$, or more precisely its equivalence class $[(\tilde h,0)]$, may be performed as follows.
We only need to consider the torsion part of $\Omega^\mathcal{B}_{D+1}(\pt)$. 
Let $Y$ be a representative manifold of a torsion element of $\Omega^\mathcal{B}_{D+1}(\pt)$.
Then there exists an integer $N$ and a $(D+2)$-manifold $Z$ such that its boundary 
is given by the disjoint union of $N$ copies of $Y$,
\beq
\partial Z = Y \sqcup \cdots \sqcup Y \qquad \text{($N$ copies of $Y$)}.
\eeq
Then, from the APS index theorem we get
\beq\label{eq:gA}
\tilde h(Y) = - \frac{1}{N} \index  \cD_{D+2}(Z)  \mod \bZ.
\eeq

When we consider Majorana fermions, we divide the $\eta$-invariant by $2$ and hence 
\beq
(h,\omega) = \frac{1}{2}(-\eta, \cI).
\eeq
Whenever we can impose a Majorana condition in $D$-dimensions, 
the APS index $\index \cD_{D+2}(Z) $ is always even. (See \cite{Witten:2019bou} for detailed explanations.) 
By repeating the above discussion, we get 
\beq\label{eq:gA2}
\tilde h(Y) =  - \frac{1}{2N} \index  \cD_{D+2}(Z)  \mod \bZ.
\eeq

Let us return to the case of SQFT $\sigma(X, \cT)$.
The appearance of the Dirac operators $\cD_{D+1}$ and $\cD_{D+2}$ is 
automatically achieved
just by considering the theories $\sigma(Y,\cT)$ and $\sigma(Z,\cT)$ for $(D+1)$-manifolds $Y$ and $(D+2)$-manifolds $Z$, respectively.
The supercharge of the sigma models correspond to the Dirac operator of the target space fermions.

The description of perturbative and global anomalies \eqref{eq:pA} and \eqref{eq:gA} (or \eqref{eq:gA2})
are almost directly related to the Witten index \eqref{eq:Iindex} and the torsion index \eqref{eq:deftorsion} 
discussed in Sec.~\ref{sec:torsion}. Let $\nu = \nu(\sigma(Z,\cT))$.
Each term $q^{n-\nu/24}$ in the $q$-expansion of $I_{\sigma(Z,\cT)}(q)$ and $J_{\sigma(Y,\cT)}(q)$ is basically the perturbative and global anomalies
of the fermions obtained from the corresponding summand $\cS_{n - \nu(X,\cT)/24}$ in the Hilbert space.
However, we need to take into account two corrections
and a clarification discussed below.

One correction is as follows.
When we increase the target space dimensions from $D$ to $D+2$,
we are not only increasing the zero modes of the sigma model 
(i.e. the quantum mechanical coordinates of the target space and gamma matrices),
but also excited modes of the chiral multiplets. Let us recall that when we realize a $D$-manifold $X$
as a boundary of a $(D+1)$-manifold $Y$, the region near the boundary is of the form $\bR \times X$.
When restricted to this region, the increased direction is just a free direction $\bR$.
The contribution of the extra excited modes from this direction is given by $\eta(\tau)^{-1}$ that comes from the left-moving
excited modes of bosons. This contribution can be cancelled by multiplying the results in $(D+1)$-dimensions
by $\eta(\tau)$. The same remark applies to the relation between $(D+1)$-dimensions and $(D+2)$-dimensions.
Therefore, we multiply the results of the computation in the sigma model $\sigma(Z,\cT)$ for $\dim Z = D+2$ by $\eta(\tau)^2$.
Another correction comes from the standard ghost field contribution if we are interested in actual heterotic string theories.
Its effect is given by $\eta(\tau)^2$. Therefore, in total we include the additional factor $\eta(\tau)^4$.

The other correction is about the Majorana condition. 
When the target space fermions are Majorana in the generalized sense of the real structure imposed by the
antiunitary operator $\mathsf{CPT}$, 
the worldsheet has the property that $\mathsf{m}$ of \eqref{eq:deftorsion}
is $2$. The reason is as follows. Majorana fermions (in addition to the chirality condition $(-1)^F=+1$)
are possible when the worldsheet $\mathsf{CPT}$ for Lorentz signature target spaces $X$ satisfies $\mathsf{CPT}^2=+1$
and $(-1)^F \mathsf{CPT} = \mathsf{CPT} (-1)^F$.
The operator algebra acting on the bundle
$\cS^{\sigma(X,\cT)}$ for a Lorentz signature $D$-manifold $X$ is of the form $\mathrm{Cliff}_{D-1,1} \hat{\otimes} \cA'$, where
$\mathrm{Cliff}_{D-1,1}$ is the Clifford algebra with the metric signature $(D-1,1)$, and $\cA'$ is the algebra of other modes.
The $\mathsf{CPT}$ gives a real structure for the representation of this algebra on $\cS^{\sigma(X,\cT)}$.
On the other hand, for a Euclidean signature $(D+2)$-manifold $Z$, the algebra is of the form $\mathrm{Cliff}_{D+2,0} \hat{\otimes} \cA''$.
The difference between $\cA'$ and $\cA''$ does not affect the real structure, and the difference between $\mathrm{Cliff}_{D-1,1}$ and $\mathrm{Cliff}_{D+2,0}$
changes the property of $\mathsf{CPT}$ from $\mathsf{CPT}^2=1$ (strict real) to $\mathsf{CPT}^2=-1$ (pseudo real).
Thus we have $\mathsf{CPT}^2=-1$ for Euclidean $(D+2)$-manifolds $Z$.
This fact implies that we have $\mathsf{m}=2$ as discussed in Sec.~\ref{sec:TI}. 
This is possible if and only if $\nu(\sigma(Z,\cT)) \equiv 4 \mod 8$.

The claim that the torsion index $J_{\sigma(Y,\cT)}(q)$ (or its multiplication by $\eta(\tau)^4$)
gives global anomalies requires the following clarification. 
In the definition of the torsion index, we have divided by $\eta(\tau)^{-\nu} \mathrm{MF}[\Delta^{-1}]_{\nu/2} \otimes \bQ$.
However, this is not appropriate when we want to study global anomalies for a fixed $\cT$. Instead,
let the subset $\mathrm{MF}_\cT [\Delta^{-1}]_{\nu/2} \subset \mathrm{MF}[\Delta^{-1}]_{\nu/2}$ be defined by
\beq
\mathrm{MF}_\cT [\Delta^{-1}]_{\nu/2} = \{ \eta(\tau)^\nu I_{\sigma(Z,\cT)} ~|~\partial Z = \varnothing,\quad \nu(\sigma(Z,\cT)) = \nu \}.
\eeq
It is possible to define an analog of the torsion index \eqref{eq:deftorsion} only for theories of the form $\cY = \sigma(Y,\cT)$ with a fixed $\cT$
by restricting our attention to $\cZ$ of the form $\cZ=\sigma(Z, \cT)$ with $\partial Z = Y^{\sqcup N}$,
\beq\label{eq:deftorsion2}
J_{\sigma(Y,\cT)}(q) =  \frac{1}{N} I_{\sigma(Z,\cT)}(q)  
\mod ~   \eta(\tau)^{-\nu} \mathrm{MF}_\cT [\Delta^{-1}]_{\nu/2} \otimes \bQ  + \mathsf{m} q^{-\nu/24} \bZ((q)).
\eeq
This is the more appropriate description of global anomalies for a fixed $\cT$ by the following reason. 
The part $ \eta(\tau)^{-\nu} \mathrm{MF}_\cT [\Delta^{-1}]_{\nu/2} \otimes \bQ$ is the set of (rationalization of) 
perturbative anomalies. We consider global anomalies only when perturbative anomalies are already absent.
Thus we are only interested in a linear combination of coefficients of the $q$-expansion of $J_{\sigma(Y,\cT)}(q) $ such that 
the linear combination is zero for 
elements of $\eta(\tau)^{-\nu} \mathrm{MF}_\cT [\Delta^{-1}]_{\nu/2} \otimes \bQ$.
Thus, \eqref{eq:deftorsion2} describes global anomalies for a fixed $\cT$.
The part $ \mathsf{m} q^{-\nu/24} \bZ((q))$ is also appropriate since
evaluation of global anomalies takes values in $\bR/\bZ$ rather than $\bR$.

Although the definition of the torsion index as in \eqref{eq:deftorsion2} is appropriate for a fixed $\cT$,
we may also consider ``stronger anomalies'' as follows. 
We no longer restrict our attention to theories of the form $\sigma(Z,\cT)$ and $\sigma(Y,\cT)$.
Instead we consider general SQFTs $\cZ$, $\cY$ with $\nu(\cZ) =\nu$ and $\nu(\cY) = \nu-1$ for a fixed $\nu$.
Then we regard $I_{\cZ}(q)$ and $J_{\cY}(q)$ (multiplied by $\eta(\tau)^4$) 
as the stronger perturbative and global anomalies. 
Notice that if anomalies are absent in this stronger sense,
then anomalies are also absent in the conventional sense of considering $\sigma(Z,\cT)$ and $\sigma(Y,\cT)$ for a fixed $\cT$.
The actual heterotic strings, i.e. $\nu=-20$ with only the $q^0$-term as the physical term, are anomaly-free in the stronger sense as we will discuss below.
Also, theories with $\nu \neq -20$ are not really heterotic strings, and our interest in them
is just as SQFTs. By these motivations, we study the stronger anomalies in which we consider all possible $\cZ$ and $\cY$.
The target space meaning of the stronger anomaly-free condition is not clear,
but it is tempting to speculate the following.
In quantum gravity, we may have transitions (or bordisms) between different internal theories. In other words
there may be bordisms from $\sigma(X,\cT)$ to $\sigma(X', \cT')$ for $\cT \neq \cT'$. (See \cite{McNamara:2019rup} for related discussions.)
If we consider all such general processes, it may be better to consider all possible SQFTs
and their bordisms. 

We conclude as follows. We consider stronger anomalies in the sense of the previous paragraph.
The target space dimension $D = \dim X$ is such that $\nu(\sigma(X,\cT)) = \nu -2$ for a fixed $\nu$.
The actual heterotic strings have $\nu=-20$.
\begin{enumerate}
\item Perturbative anomalies are given by the homomorphism
\beq\label{eq:pA0}
\cZ \mapsto \eta(\tau)^4 I_{\cZ}(q)
\eeq
where $\cZ$ is an element of SQFTs with $\nu(\cZ) = \nu$, and $ I_{\cZ}(q) = \tr (-1)^F q^{H_L}\bar{q}^{H_R}$ 
is the index.\footnote{Recall that we avoid using notations of CFT because we do not assume $\sigma(X,\cT)$ to be a CFT,
so instead we defined $H_L = \frac12 (H + P)$ and $H_R = \frac12 (H - P)$
in terms of the worldsheet Hamiltonian $H$ and momentum operator $P$.}
\item Global anomalies are given by (the negative of) the homomorphism
\beq\label{eq:gA0}
\cY \mapsto \eta(\tau)^4 J_{\cY}(q) 
\eeq
where $\cY $ is an element of SQFTs with $\nu(\cY)=\nu-1$, and it is a torsion element (i.e. there exists $N \in \bZ$ such that $\cY^{\oplus N}$ is a boundary theory of some SQFT),
and $ J_{\cY}(q)$ is the torsion index 
as defined in \eqref{eq:deftorsion}. 
\item For Majorana fermions in the sense of the real structure imposed by the worldsheet $\mathsf{CPT}$, 
we divide the results by $2$. As mentioned above, the $\sm$ appearing in \eqref{eq:deftorsion}
is given by $\mathsf{m}=2$ in that case
and hence we just get $q^{-\nu/24} \bZ((q))$ after the division by $2$. 
Equivalently, we can just consider $J_{\cY}(q) $ modulo $2q^{-\nu/24} \bZ((q))$ without dividing by $2$.
\item If we are interested in actual heterotic string theories, we take the $q^0$ term.
\end{enumerate}
We need to consider all possible SQFTs $\cZ$ with $\nu(\cZ)=\nu$ for $I_{\cZ}(q)$, and $\cY$ with $\nu(\cY) = \nu-1$
for $ J_{\cY}(q) $.

\begin{figure}
\centering
\includegraphics[width=0.8\textwidth]{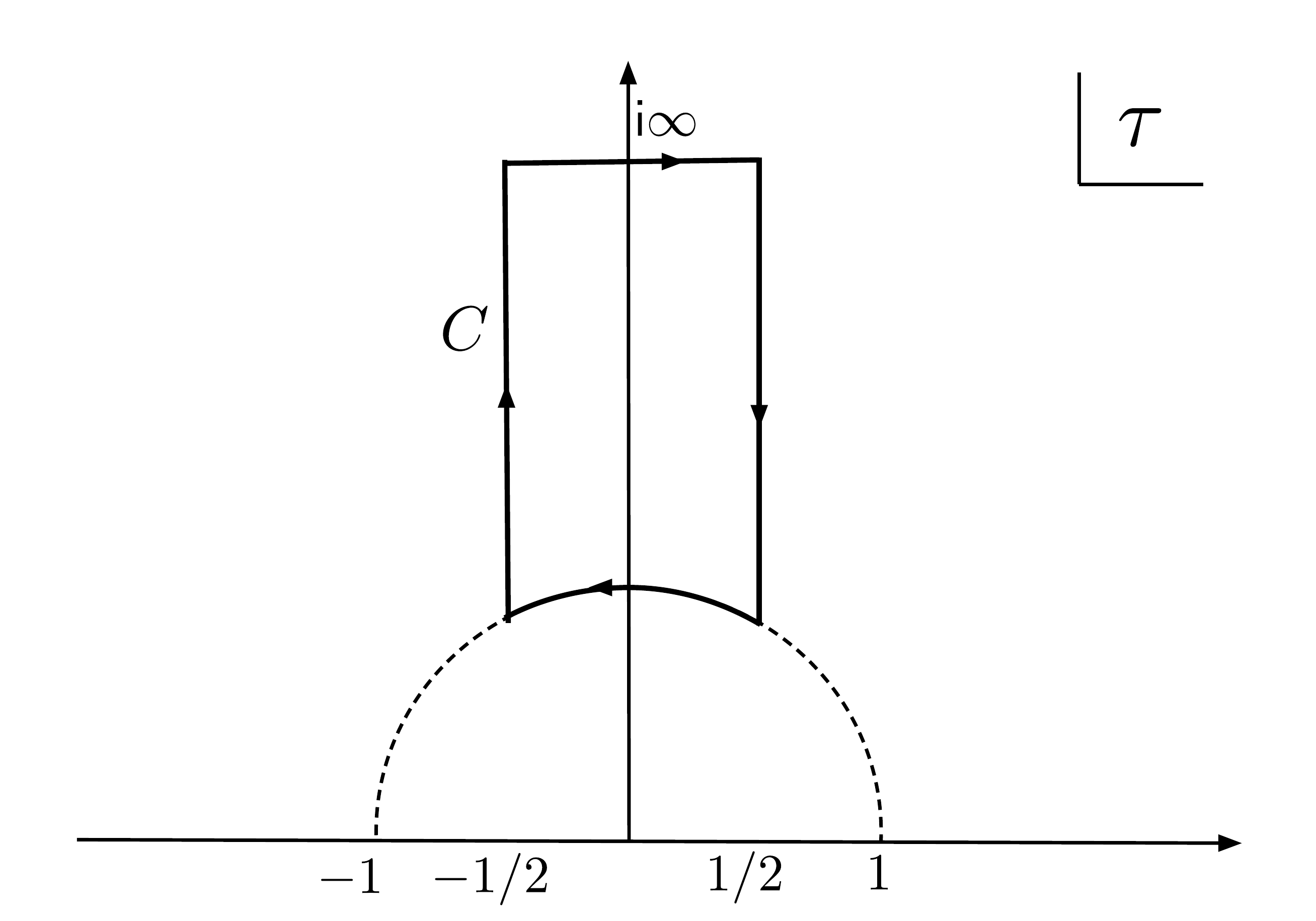}
\caption{An integration contour $C$ in the $\tau$-plane. It encloses the fundamental region of the $\SL(2,\bZ)$ action.
 \label{fig:tau}}
\end{figure}

For perturbative anomalies, the above discussion is just a rephrasing of the original discussion in \cite{Schellekens:1986xh,Lerche:1987qk,Lerche:1988np}.
Let us review the fact that perturbative anomalies are zero for the case $\nu(\cZ) = -20$. This is the case relevant for actual heterotic string theories. 
More generally, the argument applies when 
\beq
\nu(\cZ) \equiv 4 \mod 24.
\eeq
We give a proof which might give some insight into global anomalies later in Sec.~\ref{sec:globalhetero}.  
The $q^0$-term of \eqref{eq:pA0} is extracted by the integral in the $\tau$-plane,
\beq \label{eq:q0term}
\int_{ - \frac12 + \i \infty}^{+ \frac12 + \i \infty}  \eta(\tau)^4 I_{\cZ}(q)\d \tau .
\eeq
Notice that $ \eta(\tau)^4 I_{\cZ}(q)$ is a modular form of weight 2, 
so the 1-form $ \eta(\tau)^4 I_{\cZ}(q)\d \tau$ is modular invariant. 
By using this fact, we can extend the integration contour on the $\tau$-plane to the closed contour $C$ in Fig.~\ref{fig:tau}. 
It is taken to enclose the fundamental region of the $\SL(2,\bZ)$ action. Only the part $[ - \frac12 + \i \infty,   \frac12 + \i \infty]$ contributes to the integral,
because the other parts are cancelled with each other due to the modular invariance of $ \eta(\tau)^4 I_{\cZ}(q)\d \tau$.
By the Stokes theorem and holomorphy of the integrand, the integral on $C$ vanishes.
We conclude that the $q^0$ term of $ \eta(\tau)^4 I_{\cZ}(q)$ is zero when $\nu(\cZ) =4 \mod 24$.

\subsection{Examples of torsion index}\label{sec:example}
We have formulated the torsion index \eqref{eq:deftorsion} but have not yet given examples. 
Here we give some examples from sigma models.

A simple but nontrivial example is the class of sigma models discussed in \cite{Gaiotto:2019asa}. 
The target space is $S^3$, and we introduce a $B$-field such that its field strength 3-form $H$ is given by
\beq
H = H_0 + kH_1 ,
\eeq
where $k\in \bZ$ is a parameter, and $H_0$ and $H_1$ are taken as follows.
The first term $H_0$ is chosen to satisfy the anomaly cancellation condition \eqref{eq:diffcondition}. 
In the $S^3$ sigma model, there is a canonical choice for $H_0$ if 
the metric on $S^3$ is the standard metric preserving the $\O(4)$ symmetry.
The APS $\eta$-invariant on such $\O(4)$-symmetric $S^3$ is zero. 
Then we can just take $H_0=0$ to satisfy the anomaly cancellation condition. 
If the metric is not $\O(4)$ symmetric, we need to deform $H_0$ so that its integral on $S^3$ coincides with the APS $\eta$-invariant.  
In the following we just consider $\O(4)$-symmetric $S^3$ and take $H_0=0$.
The second term $H_1$ is a closed 3-form $\d H_1=0$ such that
\beq
\int_{S^3}H_1=1.
\eeq
We can add $kH_1$ without spoiling the anomaly cancellation condition \eqref{eq:diffcondition}. 
We denote the 3-sphere with the $B$-field flux $kH_1$ as $S^3_k$, and the corresponding sigma model as $\sigma(S^3_k)$.

Let us compute the torsion index of $\sigma(S^3_k)$. 
Let $Z$ be a 4-manifold obtained by removing 24 copies of a ball $B^4$ from a K3 surface, 
\beq
Z = \mathrm{K3}\setminus (B^4 \sqcup \cdots \sqcup B^4).
\eeq
Then we can realize $\sigma(S^3_{-1})^{\oplus 24}$ as the boundary theory of $\sigma(Z)$.
The condition that we need 24 copies of $S^3_{-1}$ comes from the following facts. 
The anomaly cancellation requires $\d H = \lambda(R)$ where $\lambda(R) = \frac12 p_1(R)$
is one-half of the first Pontryagin class represented by the Riemann curvature 2-form $R$.
The $\mathrm{K3}$ and hence the manifold $Z$ has $\int_{Z} \lambda(R) = -24$. 
This can be seen by Atiyah-Singer or signature index theorem on $\mathrm{K3}$.
Thus we have $\int_{\partial Z}H = \int_{Z}\lambda(R) = -24$. 

The APS index on $Z$ is the same as the Atiyah-Singer index on $\mathrm{K3}$.
For the purpose of demonstrating that the torsion index is nontrivial, 
let us focus on the lowest term in the expansion $I_{\sigma(Z )} = a q^{-1/6} + \cdots$. 
The coefficient $a$ of $q^{-1/6}$ is given by the APS index of the Dirac operator without coupling
to any other bundle. Thus it is $a=2$. From this fact, we see that
\beq
\eta(\tau)^4 J_{\sigma(S^3_{-1}) } &= \frac{1}{24} \eta(\tau)^4 I_{\sigma(Z)} \mod   \mathrm{MF}[\Delta^{-1}]_{2} \otimes \bQ  +2\bZ((q)) \nonumber \\
&=\left( \frac{2}{24} + \cdots \right)  \mod   \mathrm{MF}[\Delta^{-1}]_{2} \otimes \bQ  +2\bZ((q))
\eeq
where we have put $\nu=4$ and $\mathsf{m}=2$ in the definition \eqref{eq:deftorsion}. 
As discussed in the paragraph containing \eqref{eq:q0term}, the $q^0$ term of a weight 2 modular form is zero.
Thus, from this term, we get an invariant $2/24 \mod 2\bZ$. As discussed in Sec.~\ref{sec:TI}, 
this gives an obstruction to spontaneous supersymmetry breaking in the infinite volume worldsheet $\bR^2$.

For more general $k$, there exists a $4$-manifold $Z'$ such that $\partial Z = S^3_k \sqcup S^3_{-1} \sqcup \cdots \sqcup S^3_{-1}$
where there are $k$ copies of $S^3_{-1}$.\footnote{When $k<0$, ``$k$ copies of $S^3_{-1}$'' should be interpreted as
$|k|$ copies of $\overline{S^3_{-1}} = S^3_1$, which is the orientation reversal of $S^3_{-1}$.
One can always relate $J_{\cY}$ and $J_{\overline{\cY}}$ by taking $\cZ = \sigma(\bR) \otimes \cY$ whose boundary theory is $\cY \oplus \overline{\cY}$.
Then we find $J_{\overline{\cY}} = -J_{\cY}$.} 
From this fact, we see that 
\beq
\eta(\tau)^4 J_{\sigma(S^3_{k}) } = -k \eta(\tau)^4 J_{\sigma(S^3_{-1}) }    \mod   \mathrm{MF}[\Delta^{-1}]_{2} \otimes \bQ  +2\bZ((q)).
\eeq

The question about possible obstruction to spontaneous supersymmetry breaking in the class of sigma models $\sigma(S^3_{k})$ was raised in \cite{Gaiotto:2019asa}.
The aforementioned conjecture on topological modular forms $\mathrm{TMF}$
and the fact $\mathrm{TMF}^{-3}(\pt) = \bZ_{24}$ 
suggest the existence of an invariant taking values in $\bZ_{24}$ which forbids spontaneous supersymmetry breaking. 
An answer was proposed in \cite{Gaiotto:2019gef}. Although our invariant is basically the same,
our basic definition and computation are different.

We may also interpret the above result as a global anomaly of the target space fermion for $D=2$ sigma models $\sigma(X)$.
In fact, a $\bZ_{24}$ global anomaly was found in \cite{Tachikawa:2021mvw} which is described as follows. 
First, recall that a $D=2$ Majorana-Weyl fermion has an anomaly given by
\beq
(h,\omega) = ( -\frac12 \eta, -\frac{1}{24}\lambda(R) ) \in \widehat {(\IO^{\cB})}{}^{4}(\pt).
\eeq
The differential form part $-\frac{1}{24}\lambda(R) $ can be cancelled by adding a counterterm $-(h_\cJ, \d \cJ)$ as in \eqref{eq:Jcounter},
where we take $\cJ = -\frac{1}{24} H$. This cancellation is possible because of the condition $\d H = \lambda(R)$.
Then we get a target space global anomaly
\beq\label{eq:z24anom}
(\tilde h, 0)= (-\frac12 \eta, -\frac{1}{24}\lambda(R)  ) -(h_{\cJ}, \d \cJ) = (-\frac12 \eta + \frac{1}{24}H, 0).
\eeq
Evaluating it on $S^3_k$, we get 
\beq
\tilde h (S^3_k) = \frac{k}{24} \mod \bZ.
\eeq
This is the $q^0$-term of $-\frac{1}{2}\eta(\tau)^4 J_{\sigma(S^3_{k}) } $.
The $D=2$ sigma models $\sigma(X)$ without an internal theory $\cT$ is a ``wrong heterotic string theory'' 
and hence the target space theory can have anomalies.

Let us also briefly comment on the target space $T^n$. 
We set $H=0$, and take the spin structure in each direction to be the periodic (non-bounding, Ramond) spin structure. 
The APS index theorem on $\bR \times T^n$
(or the direct computation of the APS $\eta$-invariant on $T^n$) gives the following results. For $n=1$ we get
\beq
J_{\sigma(S^1)} = \left( \frac{1}{2} q^{-2/24} + \cdots \right)  \mod   q^{-2/24} \bZ((q)) 
\eeq
where we have used $ \mathrm{MF}[\Delta^{-1}]_{1} =0 $. For $n=3$ we get 
\beq
J_{\sigma(T^3)} = \left(1\cdot q^{-4/24} + \cdots \right)   \mod   \eta(\tau)^{-4} \mathrm{MF}[\Delta^{-1}]_{2} \otimes \bQ  +2q^{-4/24} \bZ((q)) .
\eeq
Both of them are nonzero and have order 2. The nontriviality of $\sigma(S^1)$ is also seen by the mod 2 index of the Hilbert space.
Our torsion index is not directly defined for $\sigma(T^2)$ since its pure gravitational anomaly $\nu(\sigma(T^2)) =2$ is even.
However, the nontrivial torsion index of $\sigma(T^3) = \sigma(T^2) \otimes \sigma(S^1) $ implies that $\sigma(T^2)$ must also be nontrivial.

\subsection{Global anomalies of heterotic string theories}\label{sec:globalhetero}

Let us come back to the case $\nu =-20$ and hence $\nu(\cY)=-21$.
We already know that perturbative anomalies
are absent. Let us consider global anomalies. 

Assume that the conjecture~\cite{Stolz:2004,Stolz:2011zj}  on $\mathrm{TMF}$ is true.
We have $\mathrm{TMF}^{21}(\pt)=0$. (The list of $\mathrm{TMF}^{\bullet}(\pt)$ is reproduced in an appendix of \cite{Tachikawa:2021mby}.) 
This fact implies that all SQFTs with the pure gravitational anomaly $\nu=-21$ 
can be deformed to a theory with spontaneous supersymmetry breaking.
Then the torsion index of any theory with $\nu=-21$ must be zero because the torsion index gives obstruction 
to supersymmetry breaking in the infinite volume.
We conclude that there is no global anomaly in the target space of actual heterotic string theories.

Reduction of heterotic string global anomalies to $\mathrm{TMF}^{21}(\pt)=0$ has been originally discussed 
by different (although related) mathematical arguments in \cite{Tachikawa:2021mby}. Our discussion
is more field-theoretical, and avoids some assumptions made there.\footnote{
On the other hand, we need to assume that various field-theoretical constructions discussed in this paper are well-defined.
In particular, we assume that theories like $\sigma(Z, \cT)$ can be UV-completed, at least for manifolds $Z$ which are
necessary for the purpose of computation of anomalies. We remark that $\sigma(Z, \cT)$ itself need not be UV-complete.
Only the existence of a UV-completion is assumed. 
}

It would be desirable if we could prove the vanishing of the torsion index more directly without
using the conjecture on $\mathrm{TMF}$.
A naive application of the argument for perturbative anomalies fails in the following way.
Let $\cZ$ be an SQFT with $\nu(\cZ) = -20$ whose boundary is given by $N$ copies of $\cY$ so that we have
\beq
J_{\cY}(q) = \frac{1}{N} I_{\cZ}(q)  .
\eeq
Let us denote the partition function $\tr(-1)^F q^{H_L}\bar{q}^{H_R}$ of the theory $\cZ$
as $Z(\cZ)$. We assume that this partition function transforms as expected under modular transformations.
 For compact $\cZ$ (i.e. $\cY =\varnothing $), we have $I_{\cZ}(q) = Z(\cZ)$.
The 1-form $\eta(\tau)^4Z(\cZ) \d \tau$ is modular invariant, so we have
\beq
\int_{ - \frac12 + \i \infty}^{  \frac12 + \i \infty}  \eta(\tau)^4Z(\cZ) \d \tau = \int_C \eta(\tau)^4Z(\cZ) \d \tau 
\eeq
where $C$ is the contour shown in Fig.~\ref{fig:tau}. By using the Stokes theorem, we get
\beq\label{eq:mock1}
\int_F \eta(\tau)^4 \frac{\partial Z(\cZ)}{\partial \bar{\tau}}  \d \tau \wedge \d \bar{\tau}
\eeq
where $F$ is the fundamental region of the $\SL(2,\bZ)$ action on the $\tau$-plane. 
If we had $I_{\cZ}(q) = Z(\cZ)$ and in particular if $Z(\cZ)$ were holomorphic, 
we would have finished the proof of the absence of global anomalies as in the case of perturbative anomalies. However, 
in general $Z(\cZ)$ is not a holomorphic function of $\tau$.  
In fact, we have already seen examples in Sec.~\ref{sec:example} 
for $\nu(\cZ )= 4 $ in which global anomalies are present, so this proof should fail. 

The derivative of $ Z(\cZ)$ by $\bar{\tau}$ is given by the one-point function
of the supersymmetry current in the boundary theory~\cite{Gaiotto:2019gef,Dabholkar:2020fde}. Let $\partial \cZ$ be the boundary theory.
The derivative is given by~\footnote{Without conformal invariance, we need to change not only $\bar{\tau}$ but also the volume of the worldsheet $T^2$ 
in an appropriate way for the following formula to be correct. 
We suppress the dependence on the worldsheet volume in the following discussion. 
Also,the formula in \cite{Gaiotto:2019gef,Dabholkar:2020fde} used the one-point function of the supersymmetry current rather than 
the supercharge $Q$, but the difference is just a constant factor.
}
\beq
\frac{\partial Z(\cZ)}{\partial \bar{\tau}} = \i C \frac{1}{\sqrt{\tau_2} \eta(\tau) } \langle Q\rangle_{\partial \cZ}
\eeq
where $\tau = \tau_1 + \i \tau_2$, $Q$ is the supercharge, and $C$ is a constant. 

To get some insight, let us neglect all excited modes and just consider SQM rather than SQFT.
Also, instead of the fundamental region $F$, we consider the region $ \tau_1 \in [-1/2, 1/2]$ and $ \tau_2 >0$. 
Then the role of the integration over $ \tau_1 $ is to project the states to $P=0$. Then we reduce $q^{H_L} \bar{q}^{H_R}  \to \exp( - 4\pi  \tau_2 Q^2)$.
We also neglect $\eta(\tau)$ and set $\eta(\tau) \to 1$. Thus, instead of \eqref{eq:mock1} we consider
\beq
\int_0^\infty \d \tau_2 C \frac{1}{\sqrt{\tau_2}} \tr Q  \exp( - 4\pi  \tau_2 Q^2) = \frac{1}{2} C \tr \left( \frac{Q}{|Q|} \right)
\eeq
where the trace is taken in the boundary theory $\partial \cZ$.
Recall that for sigma models $Q$ is given by the Dirac operator, $Q = \i \slashed{D}$. 
Thus, $\frac{1}{2} \tr (Q/|Q|)$ is a formal expression for the APS $\eta$-invariant.\footnote{
See also \cite{Dabholkar:2019nnc} for discussions of the APS index theorem in SQM.}
In SQFT rather than SQM, the integration region is actually $F$ rather than $[-1/2, +1/2] \times (0,\infty)$,
and there are also other contributions such as the 3-form $H$. For instance, we may get
a term in \eqref{eq:z24anom} proportional to $H$ from a term $H_{IJK}\psi^I\psi^J\psi^K$ in $Q$ as in \cite{Dabholkar:2020fde}.
In any case, the basic lesson is that the proof of vanishing of \eqref{eq:mock1} fails by the quantities that are 
related to global anomalies.
Therefore, it seems not straightforward to extend the proof of the vanishing of perturbative anomalies to the case of global anomalies. 

To illustrate the difficulty a little more, we note that there is a theory $\cT$ with the pure gravitational anomaly $\nu(\cT) = -24$
such that its partition function or the index is given by $I_\cT(q) = \tr (-1)^F q^{H_L}\bar{q}^{H_R} = 24$.\footnote{For example, such a theory
is constructed by taking 24 copies of fermi multiplets and gauging its axial $\bZ_2$ symmetry \cite{Gukov:2018iiq}. The $\bZ_2$
is anomaly free if the number of Majorana-Wely fermions is a multiple of 8 \cite{Witten:1985mj,Tachikawa:2018njr,Witten:2019bou}. }
Let us also take $Y = S^3_{-1}$ and $Z=\mathrm{K3} \setminus (B^4 \sqcup \cdots \sqcup B^4)$ that are considered in Sec.~\ref{sec:example}.
Then, $\cZ = \sigma(Z) \otimes \cT$ has the pure gravitational anomaly $\nu(\sigma(Z) \otimes \cT) = -20$.
The torsion index $J_{\sigma(Y) \otimes \cT}(q)$ is zero, but $\frac{1}{24} I_{\sigma(Z)\otimes \cT}(q)$ is 
nonzero before dividing by $\mathrm{m} q^{-\nu}\bZ((q))$. Thus we cannot hope that $ I_{\cZ}(q)$ itself is exactly zero.
We really need to take into account the fact that global anomalies are torsion.

\section*{Acknowledgements}
The author thanks Yuji Tachikawa for helpful comment on the early version of the manuscript,
and Joseph Davighi for pointing out a typo.
The work of KY is supported in part by JST FOREST Program (Grant Number JPMJFR2030, Japan), 
MEXT-JSPS Grant-in-Aid for Transformative Research Areas (A) ”Extreme Universe” (No. 21H05188),
and JSPS KAKENHI (17K14265).

\bibliographystyle{ytphys}
\bibliography{ref}

\end{document}